\documentclass[sigconf]{acmart}

\PassOptionsToPackage{protrusion, expansion}{microtype} 

\usepackage{amsmath, amsthm}
    
\usepackage{graphicx, xcolor} 
\usepackage{subfigure}
\usepackage{hyperref, url}
\usepackage{adjustbox}
\usepackage{booktabs, multirow}
\usepackage{caption}
\usepackage{enumitem}
\usepackage{lipsum}
\usepackage{setspace}
\usepackage{algorithm, algpseudocode}
    \definecolor{BLUE}{rgb}{.0, .2, .6}
    \definecolor{BLUEalt}{HTML}{1e50a2}
    \definecolor{RED}{HTML}{c9171e}
    \algrenewcommand{\alglinenumber}[1]{{\scriptsize\bfseries\ttfamily\color{RED}#1}}

\usepackage{footnote}
\makesavenoteenv{tabular}
\makesavenoteenv{table}

\usepackage[zerostyle=d]{newtxtt}

\copyrightyear{2021}
\acmYear{2021}
\setcopyright{rightsretained}
\acmConference[HPDC '21]{Proceedings of the 30th International
Symposium on High-Performance Parallel and Distributed
Computing}{June 21--25, 2021}{Virtual Event, Sweden}
\acmBooktitle{Proceedings of the 30th International Symposium on
High-Performance Parallel and Distributed Computing (HPDC '21), June
21--25, 2021, Virtual Event,
Sweden}\acmDOI{10.1145/3431379.3460653}
\acmISBN{978-1-4503-8217-5/21/06}

\begin{document}

\title{Adaptive Configuration of In Situ Lossy Compression for Cosmology Simulations via Fine-Grained Rate-Quality Modeling}

\settopmatter{authorsperrow=3}

\newcommand{\AFFIL}[4]{%
    \affiliation{%
        \institution{\small #1}
        \city{#2}\state{#3}\country{#4}
    }
    }

\author{Sian Jin}{\AFFIL{Washington State University}{Pullman}{WA}{USA}}
\email{sian.jin@wsu.edu}

\author{Jesus Pulido}{\AFFIL{Los Alamos National Laboratory}{Los Alamos}{NM}{USA}}
\email{pulido@lanl.gov}

\author{Pascal Grosset}{\AFFIL{Los Alamos National Laboratory}{Los Alamos}{NM}{USA}}
\email{pascalgrosset@lanl.gov}

\author{Jiannan Tian}{\AFFIL{Washington State University}{Pullman}{WA}{USA}}
\email{jiannan.tian@wsu.edu}

\author{Dingwen Tao}{\AFFIL{Washington State University}{Pullman}{WA}{USA}}
\authornote{Corresponding author: Dingwen Tao, School of Electrical Engineering \& Computer Science, Washington State University, Pullman, WA 99163, USA.}
\email{dingwen.tao@wsu.edu}

\author{James Ahrens}{\AFFIL{Los Alamos National Laboratory}{Los Alamos}{NM}{USA}}
\email{ahrens@lanl.gov}

\begin{abstract}
Extreme-scale cosmological simulations have been widely used by today's researchers and scientists on leadership supercomputers. A new generation of error-bounded lossy compressors has been used in workflows to reduce storage requirements and minimize the impact of throughput limitations while saving large snapshots of high-fidelity data for post-hoc analysis. In this paper, we propose to adaptively provide compression configurations to compute partitions of cosmological simulations with newly designed post-analysis aware rate-quality modeling. The contribution is fourfold: (1) We propose a novel adaptive approach to select feasible error bounds for different partitions, showing the possibility and efficiency of adaptively configuring lossy compression for each partition individually. (2) We build models to estimate the overall loss of post-analysis result due to lossy compression and to estimate compression ratio, based on the property of each partition. (3) We develop an efficient optimization guideline to determine the best-fit configuration of error bounds combination in order to maximize the compression ratio under acceptable post-analysis quality loss. (4) Our approach introduces negligible overheads for feature extraction and error-bound optimization for each partition, enabling post-analysis-aware in situ lossy compression for cosmological simulations. Experiments show that our proposed models are highly accurate and reliable. Our fine-grained adaptive configuration approach improves the compression ratio of up to 73\% on the tested datasets with the same post-analysis distortion
with only 1\% performance overhead.
\end{abstract}

\ccsdesc[500]{Theory of computation~Data compression}

\keywords{Lossy Compression; Science Data; Cosmology; In Situ; Post Analysis}

\maketitle
\pagestyle{plain}

\setlength{\textfloatsep}{6pt}

\section{Introduction}
\label{sec:introduction}

Large-scale scientific simulations running with leadership supercomputers are essential in many science and engineering domains such as cosmology studies. Modern cosmological simulations are used by researchers and scientists to investigate new fundamental astrophysics ideas, develop and evaluate new cosmological probes, assist large-scale cosmological surveys, and investigate systematic uncertainties~\cite{heitmann2019hacc,friesen2016situ}. Historically such studies have required large simulations that are highly computation and storage intensive, which are run on leadership supercomputers.
Today's supercomputers have evolved to heterogeneity with accelerator-based architectures, in particular GPU-based high-performance computing (HPC) systems, such as the Summit system~\cite{summit} at Oak Ridge National Laboratory. To adapt to this evolution, cosmological simulation codes such as Nyx~\cite{almgren2013nyx} (an adaptive mesh cosmological simulation code) have been designed to take advantage of GPU-based HPC systems and can be efficiently scaled to simulate trillions of particles on millions of cores~\cite{almgren2013nyx}. These simulations often run on a static number of ranks, usually for the same number of compute partitions, and periodically huge amounts dump raw simulation data to the storage for future post-hoc analysis.

With the increase in scale of such simulations, 
saving all the raw data generated to disk becomes impractical due to: 1) limited storage capacity, and 2) the I/O bandwidth required to save this data to disk can create bottlenecks in the simulation ~\cite{wan2017comprehensive,wan2017analysis,cappello2019use} . For example, one Nyx simulation with a resolution of $4096\times4096\times4096$ cells can generate up to 2.8 TB of data for a single snapshot; a total of 2.8 PB of disk storage is needed assuming running the simulation 5 times with 200 snapshots dumped per simulation. 
One way to avoid this issue is to limit the volume of data that needs to be written to disk. This can be done by decimation, e.g. storing one snapshot at every other timestep during the simulation. However, even with decimation, we can still be left with a massive number of timesteps to store and the amount of data to be stored for one timestep can still overwhelm the storage capacity and I/O bandwidth of a supercomputer.

\begin{figure*}[]
    \centering
    \includegraphics[width=0.9\linewidth]{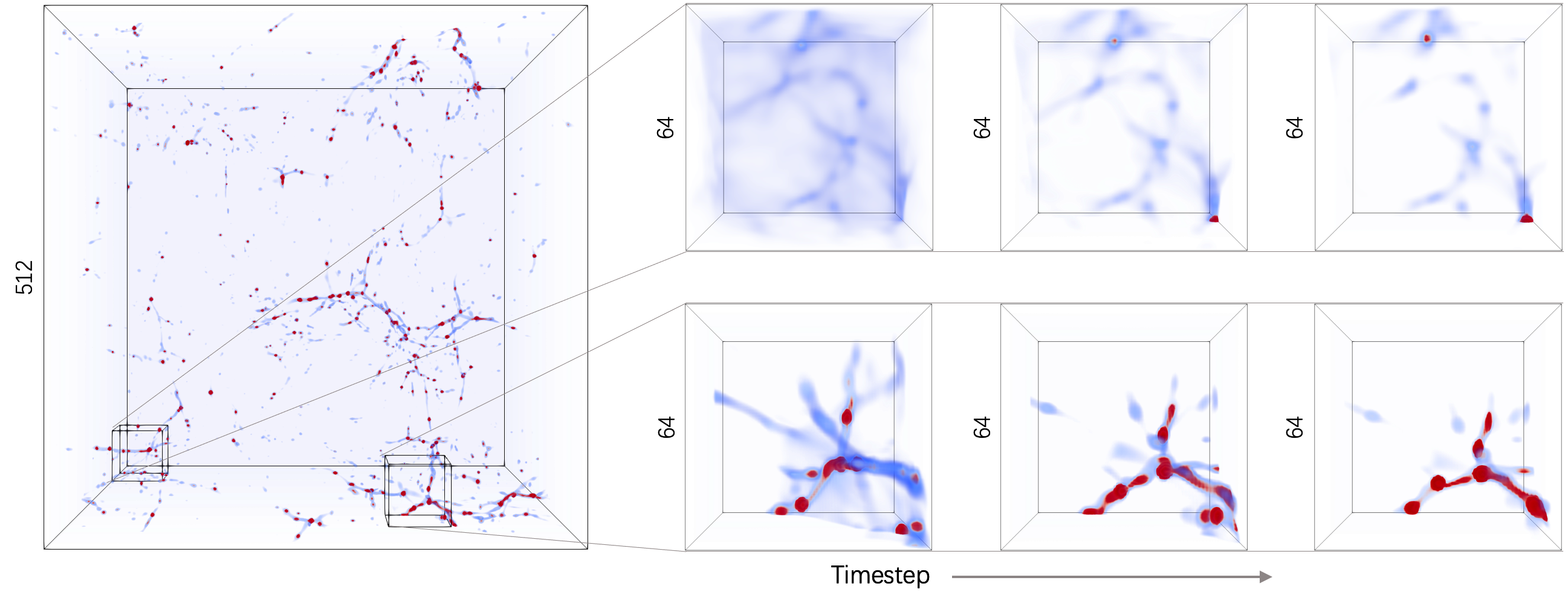}
    \vspace{-4mm}
    \caption{Left: visualization of Baryon Density in Nyx simulation under resolution of $512\times512\times512$. Right: two sample regions change through timesteps. Areas with deeper color represent for areas of higher density. Visualization magnified to enhance low values colored in blue.}
    \vspace{-4mm}
    \label{fig:fig-31-2}
\end{figure*}

A better way to address this issue is to use data compression.
While lossless compression would have been ideal, it typically only achieves a $2\times$ compression ratio~\cite{son2014data} for scientific data. On the other hand, using the new generation of error-bounded lossy compression techniques, such as SZ~\cite{tao2017significantly, di2016fast, liangerror} and ZFP~\cite{zfp}, we can achieve much higher compression ratios with minimal distortion of the data as demonstrated in many prior studies~\cite{di2016fast,tao2017significantly,zfp,liangerror,lu2018understanding,luo2019identifying,tao2018optimizing,cappello2019use,jin2020understanding,grosset2020foresight}. However, previous approaches of utilizing lossy compression for scientific datasets have always applied the same compression configuration to the entire dataset~\cite{jin2020understanding, tao2017exploration}. Yet, if we look at a visualization of baryon density in a Nyx simulation, shown in Figure~\ref{fig:fig-31-2}, we can see that not all partitions (regions) have the same amount of information. Cosmologists are typically interested in the dense regions as these would contain halos (clusters of particles) where galaxies would be formed. So, this means that the sparse regions (top row of Figure~\ref{fig:fig-31-2}) could be compressed more aggressively than the dense ones (bottom row of Figure~\ref{fig:fig-31-2}) and this would not impact the analysis done by cosmologists.

Apart from fine-grained adaptive compression, we must also be able to precisely control the compression error for domain-specific post-hoc analysis. Research has shown that general-purpose data distortion metrics, such as \textit{peak signal-to-noise ratio} (PSNR), \textit{normalized root-mean-square error}, \textit{mean relative error} (MRE), and \textit{mean square error} (MSE), on their own cannot satisfy the demand of quality for cosmological simulation post-hoc analysis~\cite{jin2020understanding,grosset2020foresight}. For example, PSNR does not tell us how the mass of a halo would be impacted after compression. One approach to finding the optimal compression configuration needed is to run a broad-spectrum analysis (try many different compression configurations and analyze the result of each) as is done by Foresight~\cite{grosset2020foresight}. However, such an approach is entirely empirical and requires a long processing time. A smarter way would be to be able to characterize each region based on a number of metrics (e.g. running an FFT analysis) which would then be used to decide which compression configuration parameters to be applied. 

In this paper, we show that adaptively compressing different regions of a simulation, based on the amount of information that they contain, allows us to maximize the compression ratio while not impacting the quality needed for post-hoc analysis. In order to determine the compression parameters to use for each region, we develop: 
(1) a theoretical error estimation model for post-hoc analysis, including both power spectrum and halo finder for cosmological simulation; 
(2) an estimation model of compression ratio for lossy compression; 
and (3) a region-wise optimization approach for error-bound combination based on the proposed models. To demonstrate the effectiveness of our approach, we compare the power spectrum and halo generated from adaptive compression to traditional static compression method and show that we get similar post-hoc analysis while getting compression ratio improvement by up to 73\%. To the best of our knowledge, this paper is the first work that systematically study the possibility and efficacy of dividing scientific simulation data in compute partitions (i.e., regions) and applying adaptive lossy compression configurations.

Regarding the overhead for in situ implementing our approach to cosmological simulation, it only requires collecting several parameters (such as mean value and number of cells weighted in a certain value range) from every region for identifying their feature density and compressibility. 
This introduces a very little overhead of only 1\% compared to compression itself since we efficiently reduce the information required for optimization. Moreover, since the data is already partitioned among MPI ranks for cosmological simulation, our approach can be perfectly integrated in situ to optimize compression configuration individually for each partition. 
Our work can also be adopted to other large-scale scientific simulations that require compression but are facing the challenge to understand the impacts of lossy compression on their domain-specific metrics.
The contributions of this work are summarized as follows:

\begin{itemize}[noitemsep, topsep=8pt, leftmargin=1.3em]
    \item We propose a novel adaptive approach to select feasible error-bound combinations for different partitions of cosmological simulation data, and show the possibility and efficiency of adaptively configuring lossy compression for each partition, instead of statically setting empirical error bound for whole data set in the beginning of the simulation.
    \item We build theoretical models to efficiently estimate (1) the overall loss of cosmological post-analysis result caused by lossy compression and (2) the compression ratio, all based on the property of each partition by collecting several representative parameters.
    \item We develop an efficient optimization guideline to determine the best-fit configuration of error-bound combinations that maximizes compression efficiency under user-defined acceptable post-hoc analysis quality distortion.
    \item Experiments demonstrate that our approach can minimize the overhead for feature extraction and error-bound optimization for each partition while providing in situ post-analysis-aware lossy compression for cosmological simulation. Our approach improves the compression ratio by up to 73\% with only about 1\% performance overhead compared to the original compression.
\end{itemize}

The rest of this paper is organized as follows. In Section~\ref{sec:background}, we discuss the background and motivation of our research. In Section~\ref{sec:analysis}, we describe our proposed modeling for cosmological simulation data post-analysis error impact and modeling for compression ratio, as well as our optimization strategy for fine-grained lossy compression. In Section~\ref{sec:evaluation}, we present the evaluation results of our proposed approach to Nyx cosmological simulation data and compare it with previous approaches. In Section~\ref{sec:conclusion}, we conclude our work and discuss our future work. 

\section{Background and Motivation}
\label{sec:background}

In this section, we present the background information about cosmology simulation and some widely used post-hoc analysis methods, as well as advanced lossy compression for scientific data. \\


\subsection{Cosmological Simulation and Analysis}
Nyx is an adaptive mesh, hydrodynamics code designed to model astrophysical reacting flows on HPC systems~\cite{almgren2013nyx,nyx}. This code models dark matter as discrete particles moving under the influence of gravity. The fluid in gas-dynamics is modeled using a finite-volume methodology on an adaptive set of 3-D Eulerian grids/mesh. The mesh structure is used to evolve both the fluid quantities and the particles via a particle-mesh method. For parallelization, Nyx uses MPI for the long-range force calculation and architecture-specific programming language for the short-range force algorithms, such as OpenMP and CUDA. Nyx data uses multiple 3-D arrays to represent field information in grid structure.  According to prior studies~\cite{nyx,habib2016hacc}, it can run up to millions of cores in the leadership supercomputers in the United States, such as Summit~\cite{summit}. 
In this paper, we use Nyx simulation data that contains 6 fields: Baryon Density, Dark matter density, Temperature, Velocity x, y, and z.

As mentioned earlier, traditional evaluation metrics (such as MSE, PSNR) cannot inform us on the data quality needed for post-hoc analysis~\cite{jin2020understanding}. So, we compute cosmology-specific evaluation metrics, such as Power Spectrum and halo characteristics, to determine the data quality needed for post-hoc analysis.

\paragraph{Power Spectrum}

Matter distribution in the Universe has evolved to form astrophysical structures on different physical scales, from planets to larger structures, such as superclusters, and galaxy filaments.
The two-point correlation function $\xi(r)$, which gives the excess probability of finding a galaxy at a certain distance $r$ from another galaxy, statistically describes the amount of the Universe at each physical scale.
The Fourier transform of $\xi(r)$ is called the matter power spectrum $P(k)$, where $k$ is the comoving wavenumber.
Therefore, the matter power spectrum describes how much structure exists at the different physical scales.
Observational data from ongoing sky surveys have measured the power spectrum of matter density fluctuations across several scales.
These sky surveys, along with large-scale simulations, are used to investigate problems such as determining cosmological parameters~\cite{eke2001power}.
In general, we compared the $P'(k)$ of decompressed data to the original $P(k)$ and target for acceptable distortion ratio within $1\pm0.01$ for all $k<10$.

\paragraph{Dark Matter Halos}

Dark matter halos play an important role in the formation and evolution of galaxies and consequently cosmological simulations.
Halos are over-densities in the dark matter distribution and can be identified using different algorithms; in this instance, we use the Friends-of-Friends algorithm~\cite{Davis1985}.
That is, we connect each particle to all ``friends'' within a distance, with a group of particles in one chain considered as one halo. Another concept of halo, such as Most Connected Particle, is defined as the particle within a halo with the most friends. Then, there is the Most Bound Particle, which is defined as the particle within a halo with the lowest potential. For the Nyx simulation, which is an Eulerian simulation instead of Lagrangian simulation, the Halo Finding algorithm uses density data to identify halos~\cite{friesen2016situ}.
For decompressed data, some of the information can be distorted from the original.
Information such as the density of one cell can affect the halo number detected, particularly for smaller halos. We use three matrices to reflex the Halo Finder quality of decompressed data: (1) the position of halos; (2) the halo number detected; and (3) the halo mass change of each halo. Furthermore, we preferred to preserve that information for middle and large halos over for small halos.

\subsection{Lossy Compression for Scientific Data}

Floating-point data compression has been studied for decades. There are two main categories: lossless compression and lossy compression. Lossless compressors such as FPZIP~\cite{lindstrom2006fast} and FPC~\cite{FPC} can only provide limited compression ratios (typically up to 2:1 for most scientific data) due to the significant randomness of the ending mantissa bits~\cite{son2014data}.
Lossy compression, on the other hand, can compress data with little information loss in the reconstructed data.
Compared to lossless compression, lossy compression can provide a much higher compression ratio while still maintaining useful information for scientific discoveries. 
Different lossy compressors can provide different compression modes, such as error-bounded mode and fixed-rate mode. 
Error-bounded mode requires users to set an error bound, such as absolute error bound or point-wise relative error bound. The compressor ensures the differences between the original data and the reconstructed data do not exceed the user-set error bound.
Fixed-rate mode means that users can set a target bitrate, 
and the compressor guarantees the actual bitrate of the compressed data to be lower than the user-set value.

In recent years, a new generation of lossy compressors for scientific data have been proposed and developed, such as SZ~\cite{di2016fast, tao2017significantly, liangerror,tian2020cusz} and ZFP~\cite{zfp}.
SZ and ZFP were first developed for CPU architectures, and both started rolling out their GPU-based lossy compression recently.
Both SZ and ZFP teams have released the CUDA implementation of their compression~\cite{tian2020cusz,cuZFP}.
Compared to lossy compression on CPUs, GPU-based lossy compression can provide much higher throughput for both compression and decompression~\cite{jin2020understanding}.
Unlike traditional lossy compressors such as JPEG \cite{wallace1992jpeg} which are designed for images (in integers), SZ and ZFP are designed to compress floating-point data and can provide a strict error-controlling scheme based on user's requirements.
In this work, we chose to use SZ instead of ZFP because the GPU version of SZ---cuSZ \cite{tian2020cusz}---provides a higher compression ratio than ZFP and offers the absolute error-bound mode that ZFP does not support (but necessary for our error control).
Specifically, SZ is a prediction-based error-bounded lossy compressor for scientific data. SZ has three main steps: (1) predict each data point's value by its neighboring data points in a multidimensional space with an adaptive predictor (using either a Lorenzo predictor \cite{ibarria2003out} or linear regression \cite{liangerror}); (2) perform an error-controlled linear-scaling quantize the difference between the real value and predicted value based on the user-set error bound, convert all floating-point values to an array of integer numbers; and (3) apply a customized Huffman coding and lossless compression to achieve a higher ratio.

Today's lossy compression techniques have been used in many HPC scientific applications for saving storage space and reducing the I/O cost of saving data \cite{gok2018pastri,wu2019full,poppick2020statistical}.
In this paper we focus on utilizing SZ lossy compression for cosmological simulation with consideration of specified analysis error control. We will build a model for SZ lossy compression error and provide a theoretical support for error propagation in post-hoc analysis. Note that our study can be also applied to other lossy compressors with modifications on compression error modeling (more will be detailed in Section~\ref{sec:analysis}).
\section{Design Methodology}
\label{sec:analysis}

In this section, we introduce the concept of optimizing compression configuration for different partitions in a given dataset and the necessary theoretical analysis as well as models needed. 
We describe in detail our theoretical analysis based on our hypothesis and provide an experimental evaluation to support our models on post-hoc analysis error impact and compression ratio. 
Lastly, we propose an in situ approach for cosmological simulation with an adaptive compression configuration for each data partition based on our analysis and model.


\subsection{Adaptive Compression Configuration}
\label{sec:3.1}

Our main goal is to provide a higher compression ratio while maintaining the same post-hoc analysis quality or provide a higher post-hoc analysis quality while maintaining the same compression ratio. We do so by applying our optimized compression configuration individually to each partition of a dataset, compared to traditionally one configuration for the entire dataset.

\begin{figure}[]
    \centering
    \includegraphics[width=0.9\linewidth]{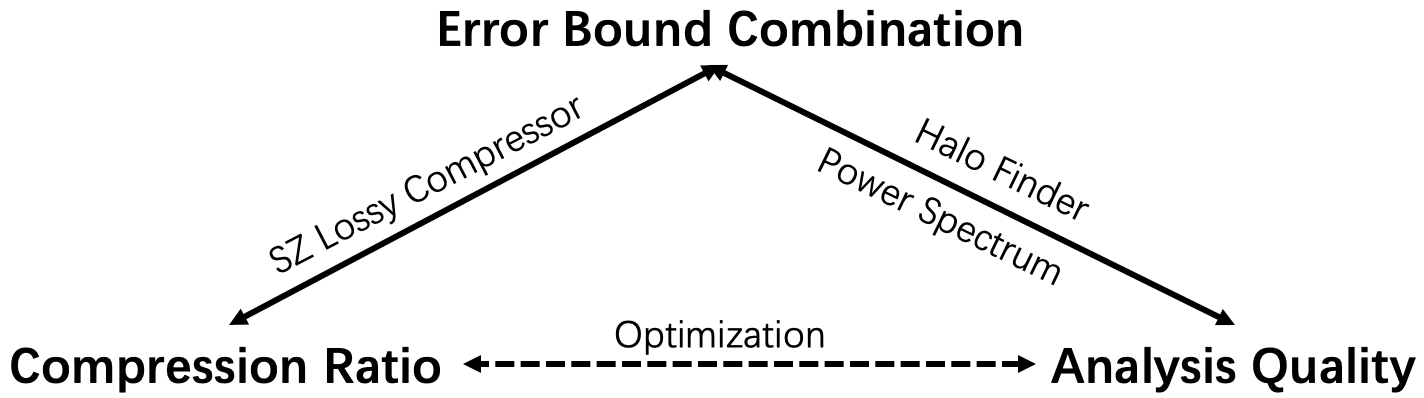}
    \vspace{-2mm}
    \caption{Quality-ratio optimization by modeling error impact on post-hoc analysis and compression ratio, based on error-bound combination of all data partitions.}
    \label{fig:fig-31-1}
\end{figure}

To achieve this, we need to determine the relationship among compression configurations, post-hoc analysis quality, and compression ratio. Thus, we introduce two types of model: (1) error-impact modeling for cosmological post-hoc analysis based on compression configurations of different partitions, and (2) compression ratio modeling based on compression configurations of different partitions. Based on these two models, we can then create an optimization strategy. 
As shown in Figure~\ref{fig:fig-31-1}, we target to balance between post-hoc analysis quality and compression ratio, when providing corresponding compression configuration for each partition. This allows us to apply different compression configurations to different partitions whereby we can improve the compression efficiency of the entire dataset by both exchanging feature preservation (e.g., preserve more feature for dense information areas) and balancing compressibility characteristic (e.g., significantly improve compression ratio by sacrificing little analysis quality for low compressibility areas). Note that we can also significantly reduce the complexity of finding the optimized solution compared to the error-and-trail baseline method when different partitions vary significantly in terms of either information density or compressibility.


As shown in Figure~\ref{fig:fig-31-2}, different partitions in Nyx data have completely different feature density and compressibility. Moreover, we can observe that a given partition can vary dramatically through different snapshots. Partitions with lower features (e.g., early timestep of sample partition upper in Figure~\ref{fig:fig-31-2}) can be almost blank in comparison to regular visualization. This means the previous compression solution of compressing all dataset with the same compression configuration is far from optimal in terms of the balance between post-hoc analysis quality and compression ratio. Note that Nyx data is naturally partitioned due to its simulation with multiple MPI ranks, which provides a suitable condition to applying an individual configuration to each data partition. Based on our proposed modeling and optimization, we adaptively adjust the compression configuration for each partition in every snapshot.


\subsection{Modeling Error Impact on Cosmology post-hoc Analysis}
\label{sec:3-sz}

\begin{figure}[]
    \centering
    \vspace{-5mm}
    \includegraphics[width=0.8\linewidth]{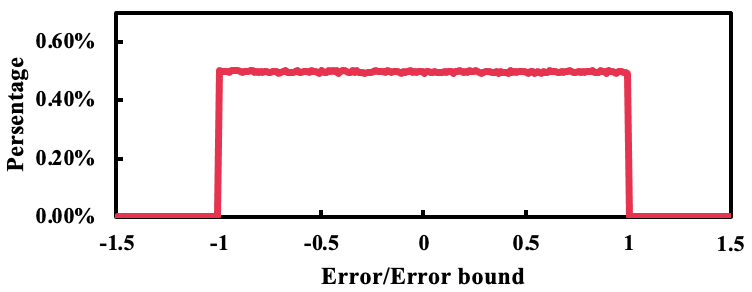}
    \vspace{-4mm}
    \caption{Error distribution of \texttt{temperature} data in one Nyx dataset compressed by SZ lossy compression with error bound of 10 and 100 bins in histogram.}
    \label{fig:fig-32-1}
\end{figure}

Power spectra and halo finder are two main post-hoc analysis metrics for Nyx cosmology simulation. We model the error impact in terms of both metrics. Note that our theoretical analysis in this section can be adopted to other post-hoc analysis with little effort. For example, our analysis for power spectra can be adopted to other FFT-based analysis.

As discussed in Section~\ref{sec:background}, SZ lossy compression is a prediction-based lossy compressor using quantization for strict error control. 
When Lorenzo predictor is used, SZ provides predicted values with an origin-prediction error in units of user-defined error bound. Such quantization causes evenly distributed error in both \verb+ABS+ mode (i.e., absolute error bound) and \verb+PW_REL+ mode (power relative error bound). More discussion in Section~\ref{sec:compression_ratio}.
Figure~\ref{fig:fig-32-1} shows the error distribution of a sample Nyx dataset on the \texttt{temperature} field with the absolute error bound of 10.0. Note that for both CPU-SZ (quantization is performed after Lorenzo prediction) and GPU-SZ (quantization is performed before Lorenzo prediction), their error distributions are the same as of uniform distribution. We also note that for some extreme cases where a high error bound is used, SZ lossy compression will introduce errors that are slightly different from the uniform distribution. We will discuss how we revise the analysis accordingly in the following sections. We use the uniform distribution as well as our revised uniform distribution to model SZ's error. Based on our evaluation, this approximation is sufficiently accurate for our optimization. 


\subsection{FFT-based Power Spectrum}
\label{sec:analysis_FFT}

As mentioned in Section~\ref{sec:background}, power spectrum analysis for Nyx cosmological simulation is based on the 3-D Fast Fourier Transform (FFT). Thus, we mainly build our error-impact model for power spectrum based on the FFT algorithm. We provide theoretical analysis on error propagation from compressor introduced error in dataset to the FFT result in terms of error distribution. We also provide an experimental evaluation to support our model.

FFT algorithm, such as Cooley-Tukey FFT algorithm~\cite{cooley1965algorithm}, utilize recursive addition and multiplication of discrete Fourier transform (DFT) matrix decomposed sparse matrix divisor to accelerate the process compared to DFT \cite{heideman1985gauss}. Moreover, FFT provides exactly the same result as DFT with lower time complexity ($O(NlogN)$ compared to $O(N^2)$). Since we introduce error to the data from lossy compression before post-hoc analysis, we can use the DFT equation instead of the more complex recursive FFT workflow to model the error.
We first start with 1-D FFT error modeling, and DFT is defined by:
\begin{equation}
    X(k) = DFT[x(n)] = \sum_{n=0}^{N-1}x(n)e^{-i\frac{2\pi}{N}nk} \quad k = 0,1,...,N-1,
    \label{equ-1}
\end{equation}
where $x(n)$ is discrete input data, $k$ is input frequency, and $N$ is the number of elements. And it can be further simplified as:

\begin{equation}
    X(k) = \sum_{n=0}^{N-1}x(n)W_{N}^{nk} \quad \text{where}\ W = e^{-i\frac{2\pi}{N}}.
    \label{equ-1.1}
\end{equation}
As of the error distribution from SZ lossy compression, as mentioned in Section~\ref{sec:3.1}, is modeled by uniform distribution as follows:
\begin{align}
    &\mathbf{eb} \sim U[-eb, eb] \nonumber\\
    &\text{where} \quad f(x) = \begin{cases}
    \frac{1}{2eb}, &-eb \leq x \leq eb, \\
    0, &\text{otherwise,}
    \end{cases}
    \label{equ-eb}
\end{align}
where $eb$ is the user-defined error bound, $\mathbf{eb}$ is the error distribution, $f(x)$ is probability density function. We consider this error as injected error to dataset and thus we can have the error model of DFT result as:
\begin{align}
    X(k)' &= \sum_{n=0}^{N-1}x(n)'\ W_{N}^{nk} = \sum_{n=0}^{N-1}(x(n)+\mathbf{eb})\ W_{N}^{nk} \nonumber\\
          &= X(k) + \sum_{n=0}^{N-1}\mathbf{eb}\ W_{N}^{nk}.
    \label{equ-2}
\end{align}
Note here $\mathbf{eb}$ is not a value but a distribution function. We can get the error distribution of DFT as:
\begin{align}
    \textbf{E}_{DFT} \sim \sum_{n=0}^{N-1}\mathbf{eb}\ W_{N}^{nk}
    \label{equ-2.1}
\end{align}

In our use case, $N$ is a relatively large number (no smaller than $512\times512\times512$ in our experiments), we can use Central Limit Theorem and know the above distribution should form into normal distribution. Now we need to find corresponding $\mu$ and $\sigma$ to define our normal error distribution. For real axis of DFT result, we can simplify Equation~\ref{equ-2.1} to:
\begin{align}
    \textbf{E}_{DFT}^{Re} &\sim \sum_{n=0}^{N-1}\mathbf{eb}\times \sin\left(\frac{2\pi nk}{N}\right).
    \label{equ-2.2}
\end{align}
Then, we can get the average individual variance $\bar{\sigma}_{individual}$ by:
\begin{align}
\bar{\sigma}_\text{individual} 
    &= \sum_{0}^{N-1} \frac{1}{N}\sqrt{\frac{\int_{-eb}^{eb} \left(x \sin\left(
        {\textstyle\frac{2\pi nk}{N}}
        \right) - 0 \right)^2\operatorname{d}\!x}{2eb}} = \sqrt{\frac{1}{6}}eb.
    \label{equ-3}
\end{align}
Note we transformed the equation with the fact that $k$ is a large number. Since the distribution of $\mathbf{eb}$ is central symmetric, the individual expected value $\mu_{individual}$ is zero. Also, for situations where the error introduced by lossy compression is not evenly distributed, we can still provide corresponding $bar{\sigma}_\text{individual}$ accordingly. Based on Central Limit Theorem, we can get the variance $\sigma$ and expected value $\mu$ of DFT error distribution are:
\begin{align}
    \sigma = \sqrt{\frac{N}{6}}eb, \quad \mu = 0,
    \label{equ-3.1}
\end{align}
where $N$ is the number of elements in given 1-D data. Similarly, we can further expand our equation to 2-D DFT results by central limit theorem, since each row in the new dimension would further perform another 1-D DFT on values with error distribution shown in Equation~\ref{equ-3.1}. And so on, we can get 3-D DFT error distribution from SZ lossy compression is:
\begin{align}
    \sigma_{3D} = \sqrt{\frac{N^3}{6}}eb, \quad \mu_{3D} = 0,
    \label{equ-3.1}
\end{align}
where $N$ is the data dimension. The same error distribution goes to $\textbf{E}_{DFT}^{Im}$ with similar analysis. As of applying various error bound to different partitions, since the element number in each partition is also considered a large number (e.g., $64^3$ when cutting $512^3$ data into 512 partitions), we can further derive our equation to:
\begin{align}
    \sigma_{3D} = \sum_{0}^{M-1}\sqrt{\frac{N^3}{6}} \frac{eb_m}{M}, \quad \mu_{3D} = 0,
    \label{equ-4}
\end{align}
where $M$ is the number of partitions. We observe that (1) the absolute error impact on FFT analysis highly relies on data size: cosmological simulations running with higher resolution are less error-tolerant regarding FFT-based post-hoc analysis; %
and (2) the FFT error distribution does not rely on high ``feature'' density areas, in other words, every value in original data has the same importance and will impact on all FFT results from error introduced by lossy compression.

\begin{figure}[]
    \centering
    \includegraphics[width=1.0\linewidth]{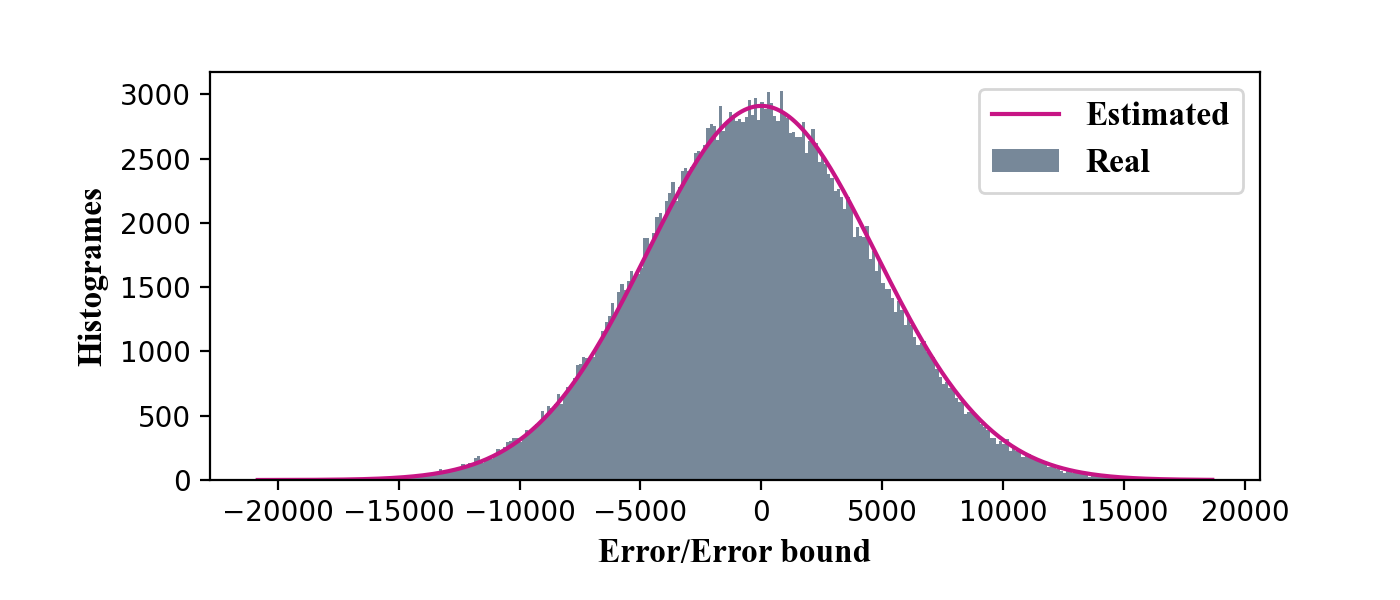}
    \vspace{-8mm}
    \caption{Comparison of real and estimated FFT error distribution based on our model using Nyx's \texttt{temperature} field.}
    \label{fig:fig-32-1}
    \vspace{-4mm}
\end{figure}

\begin{figure}[]
    \centering
    \includegraphics[width=0.55\linewidth]{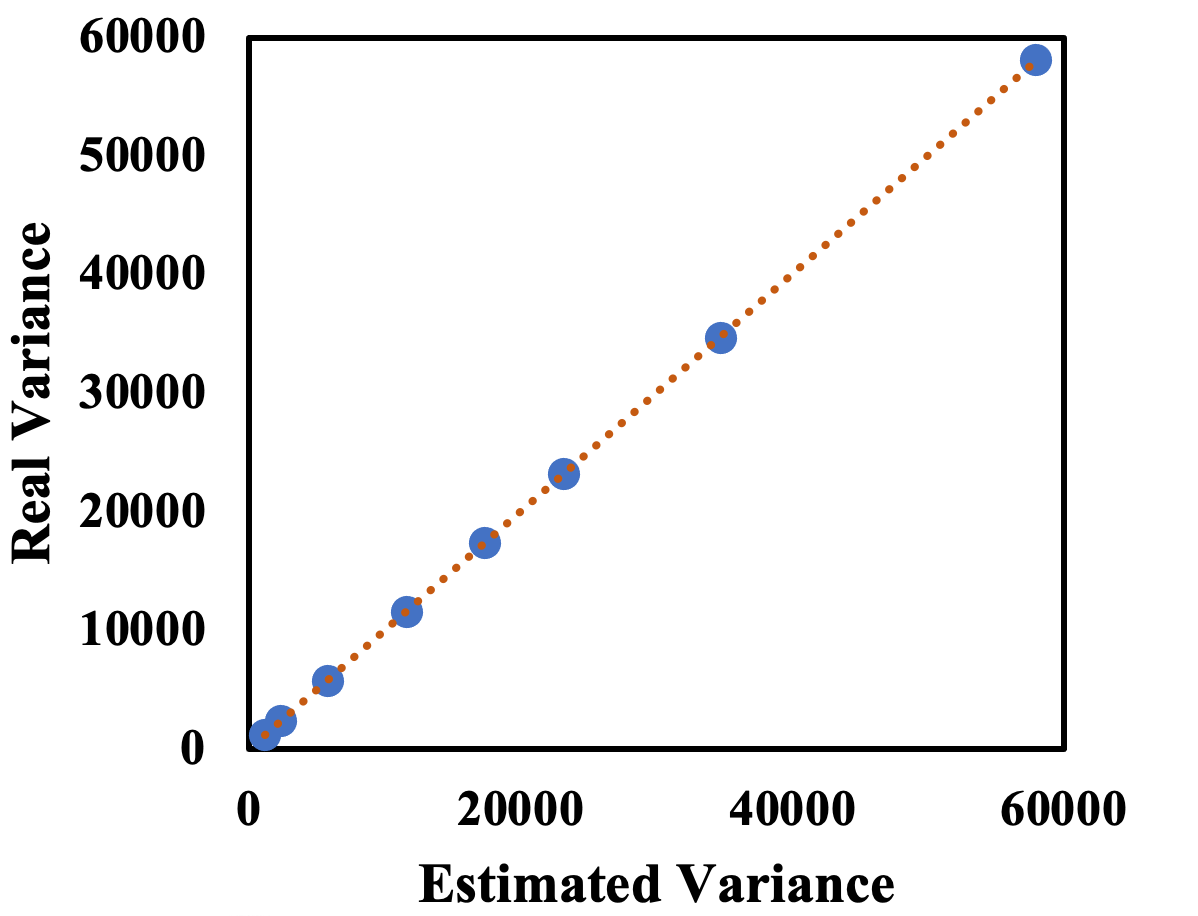}
    \vspace{-3mm}
    \caption{Comparison of real and estimated variances of FFT error based on our model using Nyx's \texttt{temperature} field.}
    \label{fig:fig-32-2}
    \vspace{-1mm}
\end{figure}

As shown in Figure~\ref{fig:fig-32-1}, we evaluate the accuracy of our model on the \texttt{temperature} field in Nyx dataset by compress the data with various compression per-partition error bound in and estimate the final impact on FFT result. The average error bound here is $1.0$, and the reason for such large error in values is because we present the error and estimation before FFT normalization, which is easy to convert by simply multiply the normalization factor. Note here x axis is normalized with error bound. Our model provides a highly reliable estimation to error impact on FFT by given compression configurations for all partitions. We also evaluate the precision of our model along with different range of error bound by applying various error bounds to partitions and collect real variance and compared to our estimation, Shown in Figure~\ref{fig:fig-32-2}. Note we no longer provide error-bounded estimation on FFT error impact, but error-bounded probability. For example, we provide a possibility of $95.45\%$ that one FFT value is within $(-2\sigma, 2\sigma)$.

\subsection{Halo Finder Analysis}

Halo finder is another post-hoc analysis for cosmological simulation such as Nyx to find halos and identify their locations and masses. Similar to our proposed model for power spectrum, we also propose a model for halo finder in order to estimate the error impact based on given compressor error in dataset, in terms of both error distribution of halo locations and halo masses.

Different from power spectrum analysis, halo finder only applied to density data field (more specifically, ``Baryon Density'') instead of all 6 fields in Nyx dataset. 
The main idea behind the halo finder algorithm is to find the areas with density higher than a given threshold $t_\text{boundary}$ and to build a tree structure across the candidates. Then, those groups that have the highest maximum larger than $t_{halo}$ are identified as ``halos''. Lastly, halo position (centroid of all grid points belong to this halo) and halo mass (cell weighted sum of all grid points belong to this halo) are recorded for each halo.

\begin{figure}[]
    \centering
    \subfigure[Original data]{
    \centering
    \includegraphics[width=0.45\linewidth]{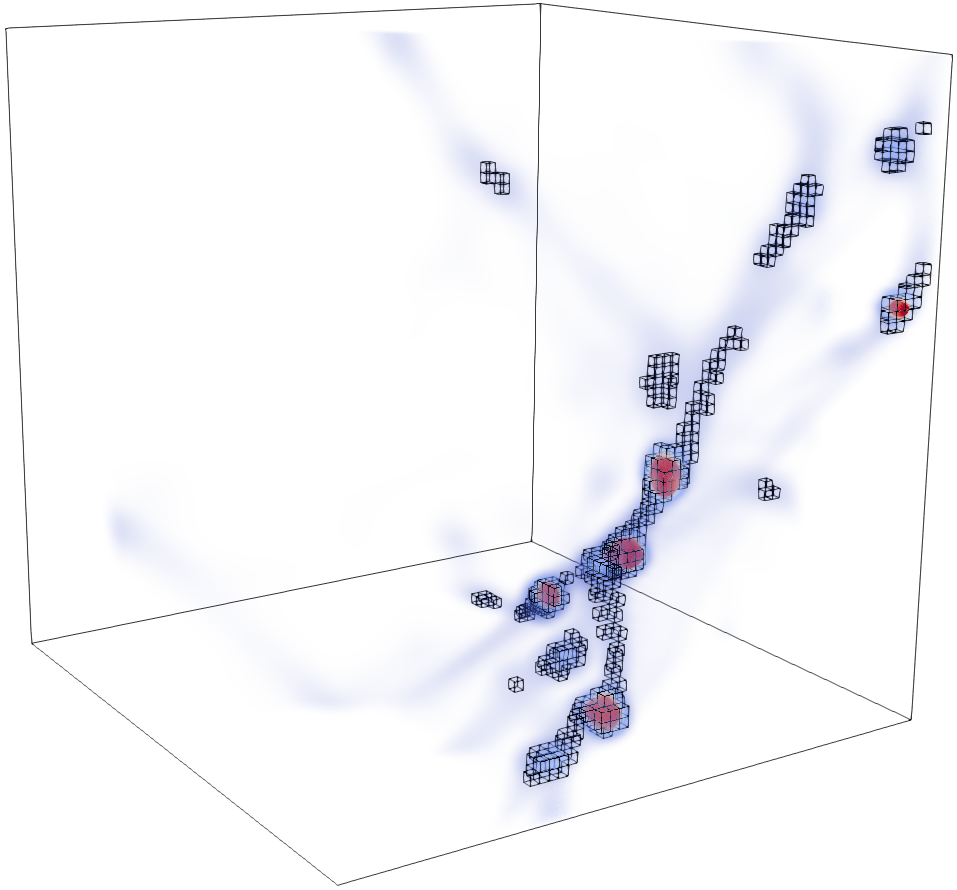}
    \label{fig:fig-33-1-ori}
    }
    \subfigure[SZ compressed data]{
    \centering
    \includegraphics[width=0.45\linewidth]{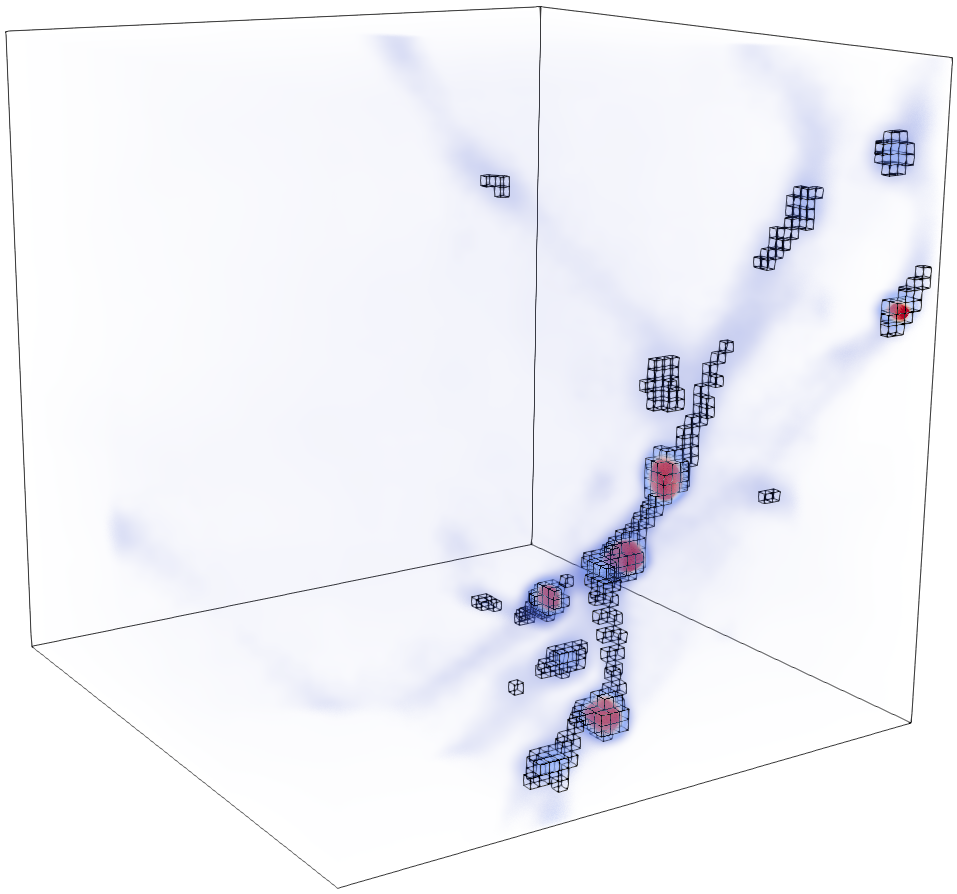}
    \label{fig:fig-33-1-sz}
    }
    \vspace{-5mm}
    \caption{Candidate cells for halo finder with  $64\times64\times64$ partitions. Areas with density higher than candidate threshold are marked in black grid.}
    \vspace{-5mm}
    \label{fig:fig-33-1}
\end{figure}

Figure~\ref{fig:fig-33-1} presents the halo candidates before and after lossy (de)compression in a given $64\times64\times64$ partition. We can observe that after lossy compression, cell candidacy changes slightly on edge areas, %
where some cells of existing halos are added or dropped. 
For demonstration, we applied a relatively high error bound of $10.0$ (normally we use error bound $eb < 1.0$), thus, more boundary cells are false-detected, and the hollow background is illuminated with reconstructed in-error-bound values visualized in blue.

\begin{figure}[]
    \centering
    \includegraphics[width=0.85\linewidth]{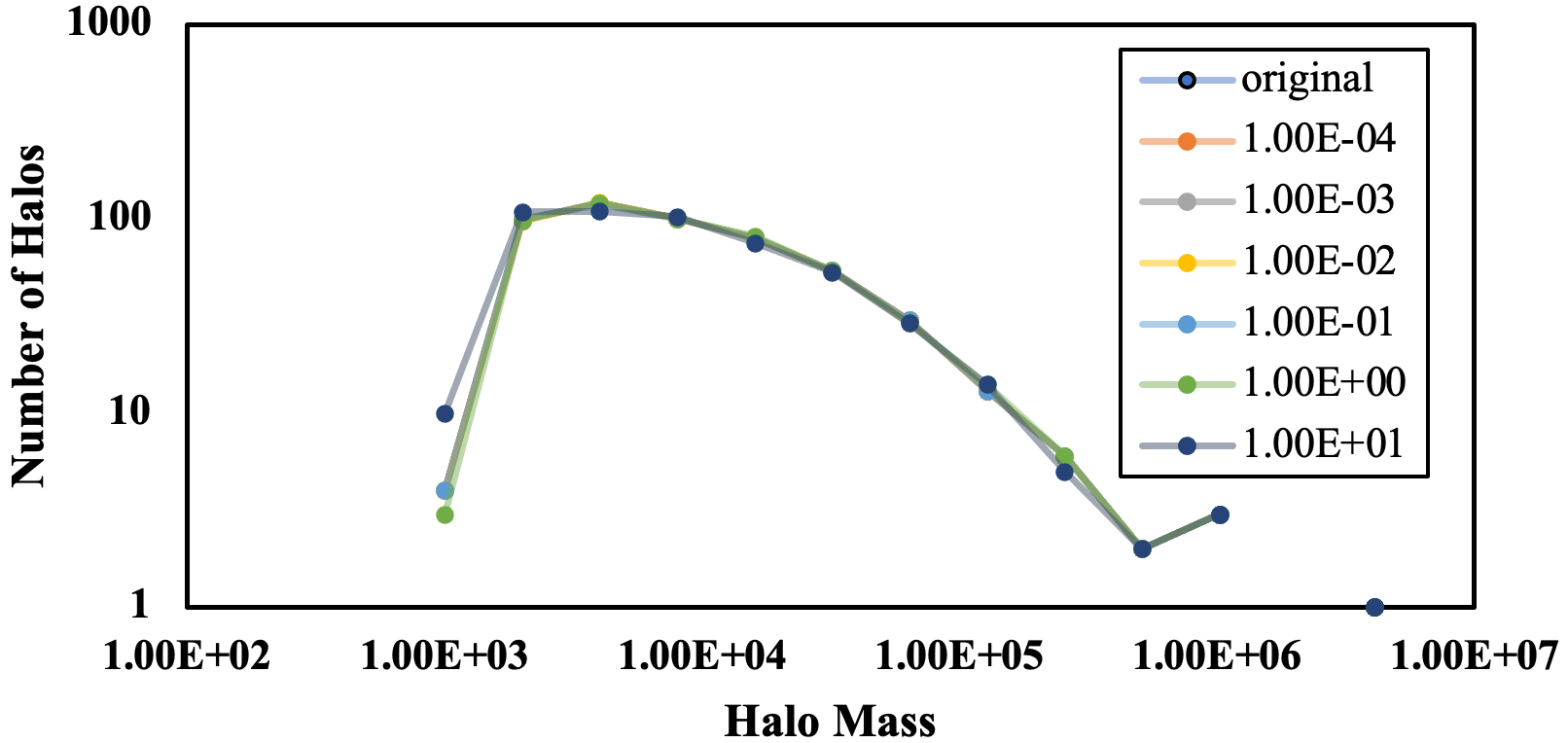}
    \vspace{-4mm}
    \caption{Comparison of halo mass distribution with different error bounds using Nyx's \texttt{baryon density} field.}
    \label{fig:fig-33-2}
    \vspace{-1mm}
\end{figure}

After compression, we observe almost no number of found halos changes, even with very high distortion from high error bound, 
as is shown in Figure~\ref{fig:fig-33-2}. 
Note that halo is further categorized into many small halos and few large halos, only the former can be false-detected under high error bound. 
In general, we intend to concentrate and preserve information for large halos based on cosmological analysis \cite{friesen2016situ}. In fact, we also observe almost no halo position change under even extreme compression during our experiment. This is reasonable since the only change we made to existing halos is minor edge distortion from introduced error by bringing edge cells up/down through the threshold. This is a very minor change in weight, compared to the central density of halos.

Compared to almost non-changing position and count after lossy compression, the mass change of halos is more suitable for error control in Nyx simulation. As shown in Table~\ref{tab:tab-1}, we analyze the mass change characteristic based on a large halo. The number of cells found for this halo increases along with the increment of the introduced error to the data as expected. However, if we target to mass difference per changed cell, we observe they fall into similar values. More specifically, it is around $88.16$, which is the set threshold for halo finder. 
It is because those edge cells are easier to switch between in- and out-of-halos given lossy compression introduced error, and they cause the mass value change of the entire cell, which is significantly larger than value change merely by lossy compression.
This means most post-hoc analysis error (from lossy compression) can be attributed to fault-detection of edge cells for halo finder.

\begin{table}[]
\ttfamily\footnotesize
\caption{Mass difference per changed cell on a large halo.}
\vspace{-3mm}
\begin{tabular}{@{}l cccc@{}}
\toprule
\bfseries Error Bound & Cells & Mass    & Mass Diff & Diff per cell \\ 
\midrule
\bfseries original    & 6023  & 3.13E+6 & -         & -             \\
\bfseries 1E-2        & 6023  & 3.13E+6 & 0         & -             \\
\bfseries 1E-1        & 6011  & 3.13E+6 & -9.8E+2   & 81.7          \\
\bfseries 1E+0        & 6038  & 3.13E+6 & 1.21E+3   & 80.7          \\
\bfseries 1E+1        & 6041  & 3.13E+6 & 1.66E+3   & 92.2          \\ 
\bottomrule
\end{tabular}
\label{tab:tab-1}
\end{table}

To further provide an estimation of mass changes given per-partition error bound, we conduct fault cell detection estimation. To start with, we divide our problem into finding fault mass detection estimation of each partition and provide overall estimation with their sum:
\begin{align}
    M_\text{fault} = t_\text{boundary}\sum_{0}^{M-1}e_{m},
    \label{equ-32.1}
\end{align}
where $M_\text{fault}$ is sum of individual halo absolute mass changes, $t_\text{boundary}$ is the threshold for halo finder, $M$ is the number of partitions, $e_{m}$ is the error in each partition. The reason is that for existing halos cell changes are addable even for halos across partitions, and for fault-detected halos, they tend to be small halos and only appear under rarely used very high error bound.

Different from FFT, which is based post-hoc analysis present in the previous section, halo finder is highly related to density as a feature of each partition for post-hoc analysis error estimation. In this case, we define the feature needs to preserve for halo finder in the given data been edge cell count around the threshold. By analyzing the value histogram of density data from Nyx simulation, we find that although histogram is not evenly distributed among the entire data range, the histogram of local values can be considered as evenly distributed. For example, if we consider error bound $eb$ as $0.1$ and halo boundary threshold $t_\text{boundary}$ as $88.16$, in which case only values within $(88.06, 88.26)$ can affect number of cells detected for halos, and the data histogram within the small value range is considered as evenly distributed. Thus, we can conclude the possibility of fault detection of given cell as:
\begin{align}
    p_\text{fault} = \frac{1}{2}\int_{t_\text{boundary}}^{t_\text{boundary}+eb}\frac{x-t_\text{boundary}}{eb}\operatorname{d}\!x = 25\%.
    \label{equ-32.2}
\end{align}
Note here similar to what we discussed for FFT-based post-hoc analysis, we can provide the corresponding $p_\text{fault}$ based on error distribution from lossy compression other than uniform distribution. Then we can provide the number of fault detected cells in the given partition by:
\begin{align}
    e_{m} = n_{bc}/4,
    \label{equ-32.3}
\end{align}
where $m$ is number of partitions for Equation~\ref{equ-32.1}, $n_{bc}$ is boundary cells with a value range of $(t_\text{boundary}-eb, t_\text{boundary}+eb)$ in the given partition. Note here we discuss the number of cells estimated been fault detected due to error introduced by lossy compression. 
Fault detected cells can switch between both in- and out-of-halos and result in expected total number of cells being the same as the original, while error forming into normal distribution similar to what we discussed in Section~\ref{sec:analysis_FFT}. However, since we focus on cell changes of individual halos and most are small halos with little edge cells, the number of cell difference can be simplified to Equation~\ref{equ-32.3}. For large halos, depending on their size, the estimated error distribution of cell count can be given by central limit theorem forms into normal distribution:
\begin{align}
    \sigma = \sqrt{\frac{n_{bc}}{3}}, \quad \mu = 0.
    \label{equ-32.3}
\end{align}

\begin{figure}[]
    \centering
    \includegraphics[width=0.77\linewidth]{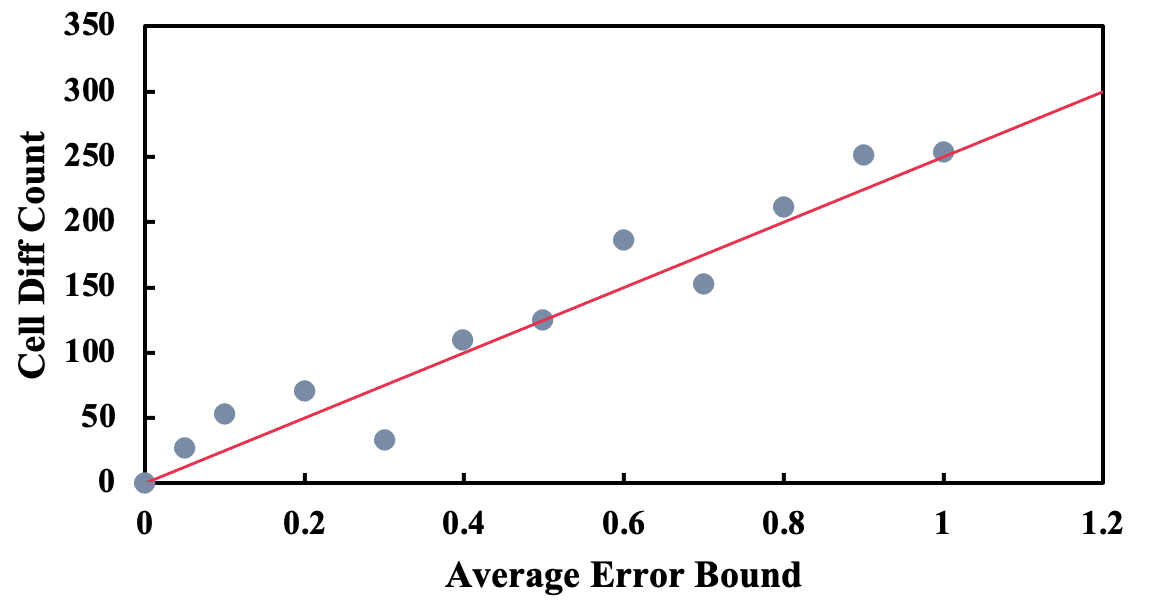}
    \vspace{-4mm}
    \caption{Number of candidate cells changed with different error bounds. Red line is the estimated cell count difference. Blue dots are the real cell count difference.}
    \label{fig:fig-33-3}
\end{figure}

We evaluate our halo finder modeling with \texttt{baryon density} data from $512\times512\times512$ Nyx simulation. Figure~\ref{fig:fig-33-3} shows our estimated cell difference count based on Equation~\ref{equ-32.1} compared to the result from applying multiple error bounds to different partitions. Our modeling provides high accuracy compared to an experimental result. Note here some of the larger differences between estimation and experiment can be due to loss/increase of small halos as well as cell difference count reduced for extremely large halos from Equation~\ref{equ-32.3}.

\subsection{Modeling Compression Ratio}
\label{sec:compression_ratio}

In this section, we build a model to estimate the overall compression ratio of the dataset based on given compression configurations of different partitions. As discussed in Section~\ref{sec:3-sz}, the error introduced by SZ lossy compression can be modeled by a uniform distribution. However, in situations of much larger error bound, the error distribution of SZ lossy compression forms into a combination of uniform distribution and normal distribution. 
This is because under high error bound settings, the Lorenzo predictor in SZ lossy compressor predicts more data-point within the error bound against the original, even without quantization. 
As mentioned in the previous theoretical analysis for post-hoc analysis, we can easily adopt our models based on revised $\sigma$ of none-even distributed error distribution. Because of this distortion of error distribution, it is extremely hard to provide a model for estimate compression ratio from error-bound combination purely with theoretical analysis. Thus, in this section, we mainly contribute to the compression-ratio modeling based on empirical analysis and provide a universal equation for all fields of data from all snapshots.

\begin{figure}[]
    \centering
    \includegraphics[width=0.95\linewidth]{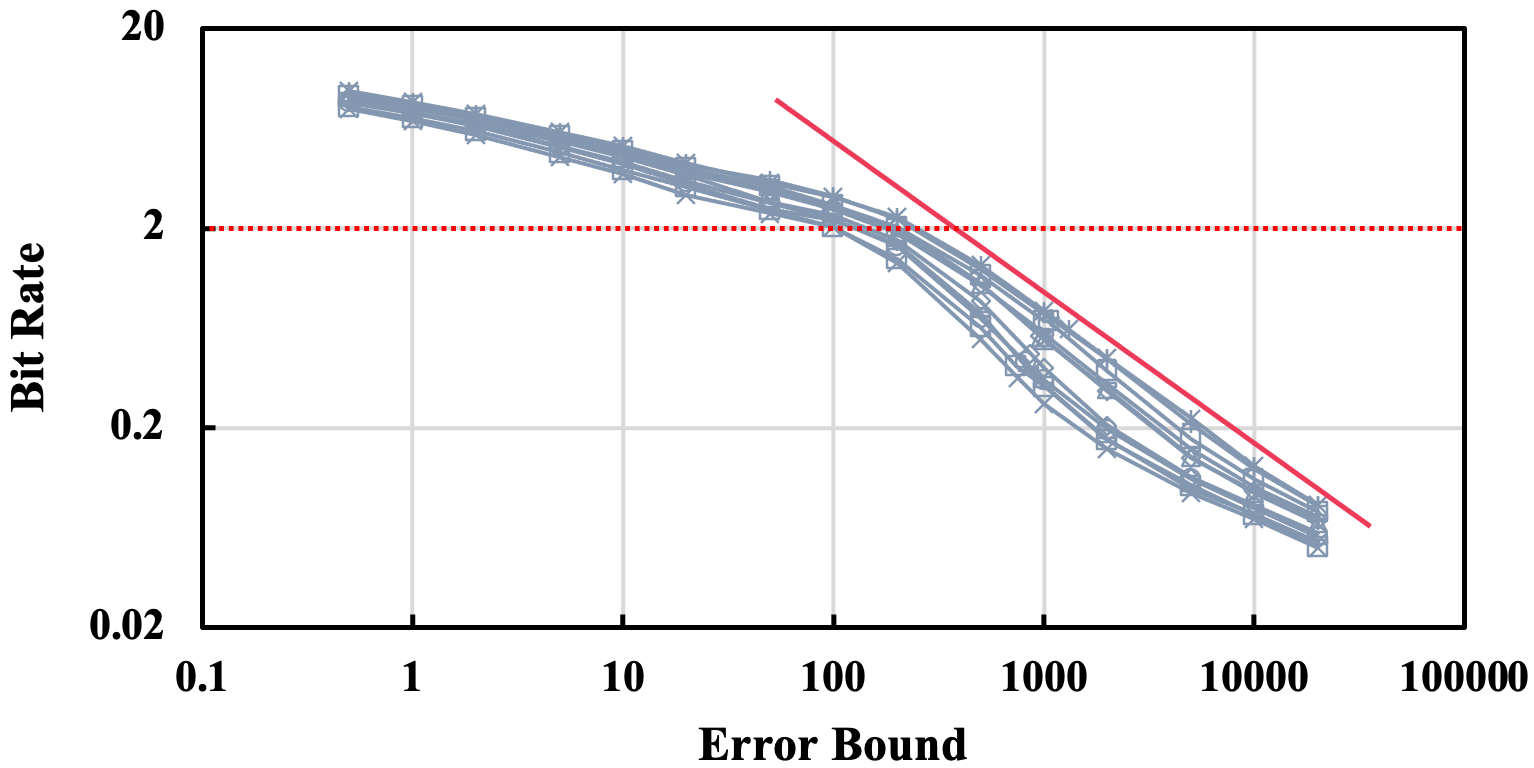}
    \vspace{-4mm}
    \caption{Bit rate with different error bounds using SZ lossy compression. Different lines represent for different partitions. 16 partitions are sampled for demonstration purpose.}
    \label{fig:fig-34-1}
    \vspace{-6mm}
\end{figure}

Figure~\ref{fig:fig-34-1} shows the bit-rate to error-bound curve by SZ lossy compression. Here bit rate represents how many bits are needed to represent a value on average, when original data is 32-bits single-precision float-point number. Two areas can be distinguished for each curve of given partition: both form into power function but areas with bit rate higher than 2 features power values closer to zero (flatter). 
This is because once the bit rate is lower than 2, a high error bound causes Lorenzo predictor to predict values within error bound (i.e. no quantization) and improves the Huffman coding efficiency, thus encourages compressibility: curves converge faster when bit rate across 2 on log scale.
\begin{figure}[]
    \centering
    \subfigure[Relative $C_1$ Estimation]{
    \centering
    \vspace{-4mm}
    \includegraphics[width=0.45\linewidth]{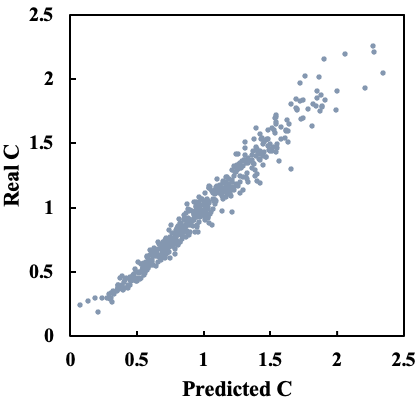}
    \vspace{-6mm}
    \label{fig:fig-34-2.1}
    }
    \subfigure[Compression Ratio Consistency]{
    \centering
    \vspace{-4mm}
    \includegraphics[width=0.45\linewidth]{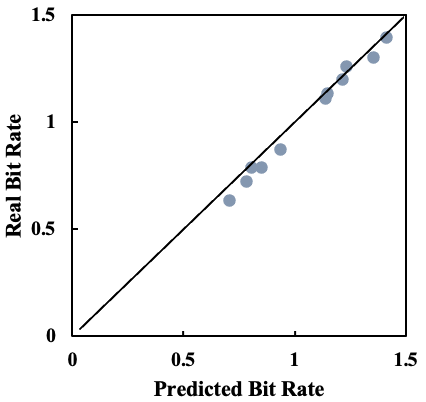}
    \vspace{-6mm}
    \label{fig:fig-34-2.2}
    }
    \vspace{-4mm}
    \caption{Accuracy on predicted $C_1$ (in Equation~\ref{equ-34.1}) and the consistency of compression ratio using SZ lossy compression.}
    \label{fig:fig-34-2}
\end{figure}
Based on our empirical studies and previous work \cite{jin2020understanding}, we consider the case where bit rate is always lower than 2, or compression ratio higher than $16\times$ for Nyx dataset (in \verb+fp32+). Also, consider our optimization strategy is only gently adjusting the error bound of different partitions, we assume our adjustment is also located within the same curve area. For partitions beyond this assumption, their impact is negligible due to model consistency on edge situation and their low percentage. Evaluation from Section~\ref{sec:evaluation} verified our assumption. Bit-rate to error-bound equation can be given by:
\begin{equation}
    B = \sum_{0}^{M-1}\frac{b_m}{M}, \quad b_m = C_meb^{c},
    \label{equ-34.1}
\end{equation}
where $B$ is overall bit rate, $M$ is the number of partitions, $b_m$ is the estimated bit rate of each partition, $C_m$ and $c$ are parameters to define position and shape of this curve. Furthermore, we can determine per-partition the basic bit-rate to error-bound power function method based on trial and error. However, we avoid this time-consuming process by performing a two-step procedure: (1) we identify that different partitions across fields and snapshots share the same power parameter $c$, thus we can select $c$ and keep using it; and (2) we use a representative parameter from a given data partition to determine $C_m$ of its bit-rate to error-bound curve, or relative compressibility of the partition. We find the entropy of partition is one of the parameters that are highly related to $C_m$, with higher entropy mapped to higher compressibility thus lower curve in Figure~\ref{fig:fig-34-1}. However, for reducing the overhead to compute parameter, we instead choose mean value as our key parameter to determine the relative compressibility of a partition.  Figure~\ref{fig:fig-34-2.1} shows that the estimated $C_m$ based on a logarithmic fitting and partitions' mean values are highly precise compared to the real $C_m$.
Lastly, unlike transform-based compressors, SZ provides consistent bit-rate to error-bound curves, as shown in Figure~\ref{fig:fig-34-2.2}. Therefore, we have high confidence to use the estimated bit rate.



\subsection{Proposed Optimization Strategy}

Until now we have built the models for error impact of post-hoc analyses as well as the model for compression ratio. Then, we optimize the compression configuration of different partitions to feature different error bound so that we can maximize the compression ratio while maintaining the estimated post-hoc analysis quality or maximizing the post-hoc analysis quality with the same estimated compression ratio. 

First, we consider power spectrum post-hoc analysis quality to compression ratio optimization. Since FFT-based post-hoc analysis does not have a correlation with individual local partition feature, meaning every value shares the same importance. The optimization strategy relies on the difference in compressibility among data partitions. Based on Equation~\ref{equ-4}, we expect a similar error impact on FFT-based analysis under the condition when the average error bound of all partitions remains the same. In this case, to improve the overall compression ratio while guaranteeing the same FFT-based post-hoc analysis quality, we utilize Equation~\ref{equ-34.1} from the previous section by estimating parameters based on the mean value of a given partition. 
Additionally, we also extract the overall mean value of the entire dataset by \verb+MPI_Allreduce+ after each partition computes their own. 
We then optimize the per-partition error bound such that their derivatives of bit-rate to error-bound curve are the same, which are the minima for Equation~\ref{equ-34.1}, while keeping the average error bound within the threshold request from Equation~\ref{equ-4}. These minima can be given by:
\begin{align}
    eb_{m} = eb_\text{avg}\cdot \exp\left\{\textstyle\frac{\ln(C_m/C_a)}{c}\right\},
    \label{equ-35.1}
\end{align}
where $eb_{m}$ is the optimized error bound of partition $m$, $eb_\text{avg}$ is the average acceptable error bound based on Equation~\ref{equ-4}, $C_a$ is the $C_m$ parameter based on the average of the mean values from Equation~\ref{equ-34.1}, $C_m$ and $c$ are each partition's parameters in Equation~\ref{equ-34.1}.

For \texttt{baryon density}, we can apply both power spectrum and halo finder post-hoc analyses. 
In this case, instead of building an optimization strategy of high complexity with three models in mind: power spectrum, halo finder, and compression ratio, we only optimize based on two of the three. Those are power-spectrum-to-ratio and halo-finder-to-ratio, from which we choose one as acceptable condition for both. 
For example, once we determine the optimized compression configuration of all partitions for power spectrum, we then evaluate if the condition is also acceptable for the halo finder. If acceptable, this combination of compression configurations is set; if unacceptable, we further apply the optimization of halo finder and ratio and then set the combination result as a boundary condition. The optimization between halo finder and compression ratio is similar to the above FFT-based post-hoc analysis optimization, but it has to change the boundary condition with Equation~\ref{equ-32.1}. In this case, we need each partition's mean value for bit-rate curve estimation as well as its number of cells with a weighted value range of $(t_\text{boundary}-eb, t_\text{boundary}+eb)$ for halo finder estimation. However, both parameters and optimizations can be calculated with little effort compared to the computationally intensive post-hoc analysis and cosmological simulation, hence introducing a very low overhead to the system with higher compression efficiency.

For both power spectrum and halo finder post-hoc analysis optimizations, we also introduce thresholds for error-bound control. This is because there may be a few partitions that may not fit our models well and produce highly unreasonable error-bound options, which can be harmful to the efficiency of our strategy. Thus, we set the highest and lowest error-bound thresholds to $4\overline{eb}$ and $\overline{eb}/4$, respectively, where $\overline{eb}$ is the average error bound over all partitions. 


\section{Experimental Evaluation}
\label{sec:evaluation}

We present the evaluation results of our framework for a fine-grained adaptive lossy compression based on our rate-quality modeling. We compare our results based on the de facto static configuration that is normally applied to datasets at the start of simulations in terms of both snapshots and multiple simulation scales. Finally, we evaluate and discuss the performance of our solution for improved compression quality.

\subsection{Experimental Setup and Dataset}

\begin{table}[]
\renewcommand*{\arraystretch}{1.3}
\centering
\ttfamily\footnotesize
\caption{Details of Nyx Dataset Used in Experiments}
\vspace{-4mm}
\newcommand\alignmiddle[2]{
\makebox[3em][r]{$(#1$}\makebox[.8em]{$,\ $}\makebox[2em][l]{$#2)$}}

\begin{tabular}{@{} c c | l c @{}}
\toprule
    Dimension
     & Size
     & Field
     & Value Range
    \\ 
\midrule
    \multirow{4}{*}{
        \begin{tabular}[c]{@{}c@{}}
            $\phantom{0}512\times\phantom{0}512\times\phantom{0}512$    \\
            $1024\times1024\times1024$ \\
            $2048\times2048\times2048$
        \end{tabular}
    }
     & \multirow{4}{*}{
        \begin{tabular}[c]{@{}r@{}}
            6.6 GB \\ 52 GB \\ 352 GB
        \end{tabular}
    }
     & Baryon Density
     & \alignmiddle{0}{10^5}
    \\ 
     &
     & Dark Matter Density
     & \alignmiddle{0}{10^4} 
    \\ 
     &
     & Temperature
     & \alignmiddle{10^2}{10^7} 
    \\ 
     &
     & Velocity
     & \alignmiddle{-10^8}{10^8} 
    \\ 
    \bottomrule
\end{tabular}

\label{tab:DataDetail}
\end{table}

We conduct our evaluation with Foresight \cite{grosset2020foresight}, an open-source toolkit used to evaluate, analyze, and visualize lossy compressors for extreme-scale cosmological simulations . We modified the toolkit so that we can gather necessary parameters for our framework to deploy adaptive lossy compression configuration to various data partitions in Nyx cosmological simulation. The details of the Nyx datasets used for evaluation are shown in Table~\ref{tab:DataDetail}. The $512\times512\times512$ Nyx dataset
is provided by the Nyx development team at Lawrence Berkeley National Laboratory \cite{k8gb-vq78-19}. It is a single-level grid structure without adaptive mesh refinement (AMR) and does not include particle data. It contains six 3-D arrays, include baryon density $(\rho_b)$, dark matter density $(\rho_d{}_m)$, temperature
$(T)$, and velocity in three directions $(v_x, v_y, v_z)$. The dataset is in the HDF5 file format~\cite{folk1999hdf5}. 
Our experiment platforms include the Cori system \cite{cori} at NERSC and the Frontera system \cite{frontera} and its subsystem Longhorn system \cite{longhorn} at TACC, of which each GPU node is equipped with 4 Nvidia Tesla V100 GPUs \cite{nv100} per node.

\subsection{Quality-Ratio Evaluation}



\begin{figure}[]
    \centering
    \includegraphics[width=0.95\linewidth]{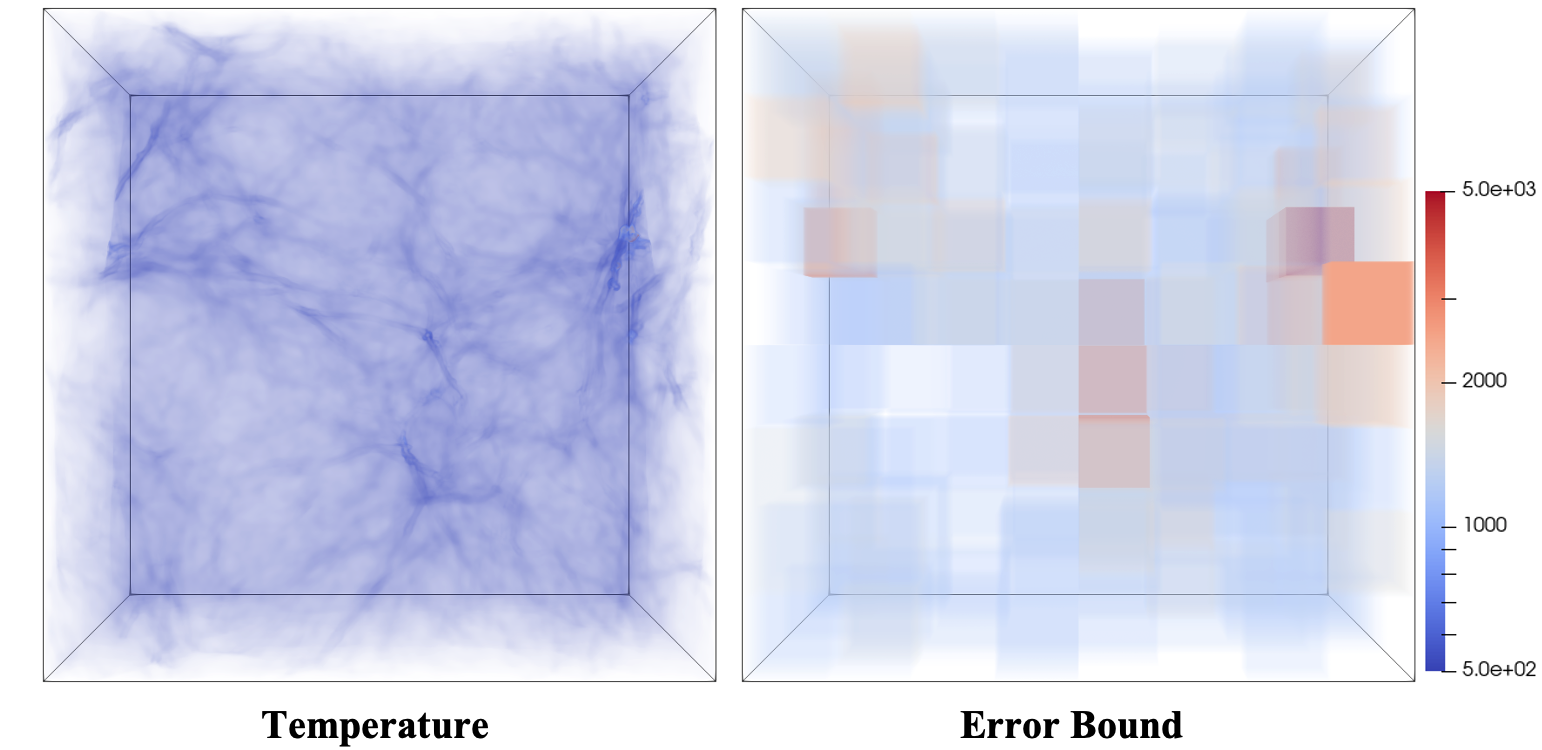}
    \vspace{-4mm}
    \caption{Fine-grained lossy compression control for different data partitions. Left: visualization of  \texttt{temperature} field in $512\times512\times512$. Right: error-bound configurations for all 512 data partitions based on our proposed method.}
    \label{fig:fig-4-1}
\end{figure}

We first experimentally demonstrate the effectiveness of our proposed solution. Figure~\ref{fig:fig-4-1} shows the visualization of error-bound of all partitions by adaptive optimization to the tested dataset. data partitions have been assigned with various error-bound, instead of the traditional method that compresses the entire dataset with a single error-bound configuration. 
Moreover, the presented \texttt{temperature} data only serve for power spectrum as post-hoc analysis, thus quality-ratio of partitions are effectively traded based on their compressibility since no features are expected to be preserved for power spectrum based on our previous analysis in Section~\ref{sec:analysis}.

\begin{figure}[]
    \centering
    \vspace{-8mm}
    \includegraphics[width=0.92\linewidth]{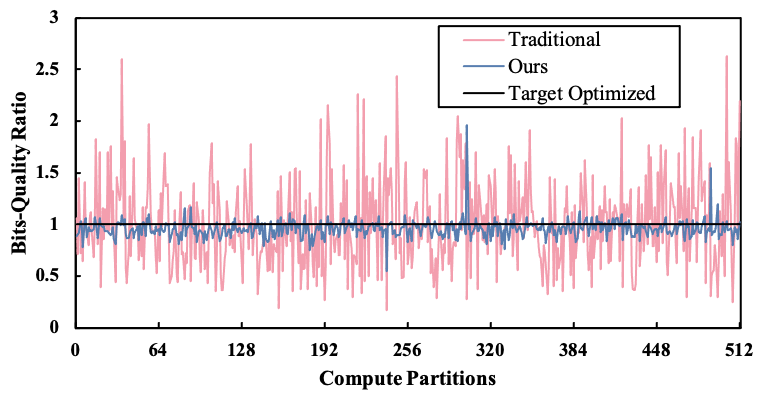}
    \vspace{-4mm}
    \caption{Comparison of bit-quality ratios using traditional and our methods on all partitions of $512\times512\times512$ Nyx's \texttt{temperature} field with 512 data partition. Y-axis is normalized to average bit-quality ratio after our optimization.}
    \label{fig:fig-4-2}
\end{figure}

In Figure~\ref{fig:fig-4-2}, we present the optimization efficiency of our optimization solution, where the bit-quality ratio is the derivative of the bit-quality curve. Traditionally, we compress every data partition with the same error-bound which results in a disorganized bit-quality ratio. This indicates the potential compression efficiency improvement by exchanging compression ratio and post-hoc analysis data quality between partitions for a higher overall compression ratio and better overall data quality. By doing so, we can significantly improve the bit-quality ratio difference of every partition, at which point every partition shares a similar balance between compression ratio and data quality as optimized.

\begin{figure}[]
    \centering
    \vspace{-2mm}
    \includegraphics[width=0.9\linewidth]{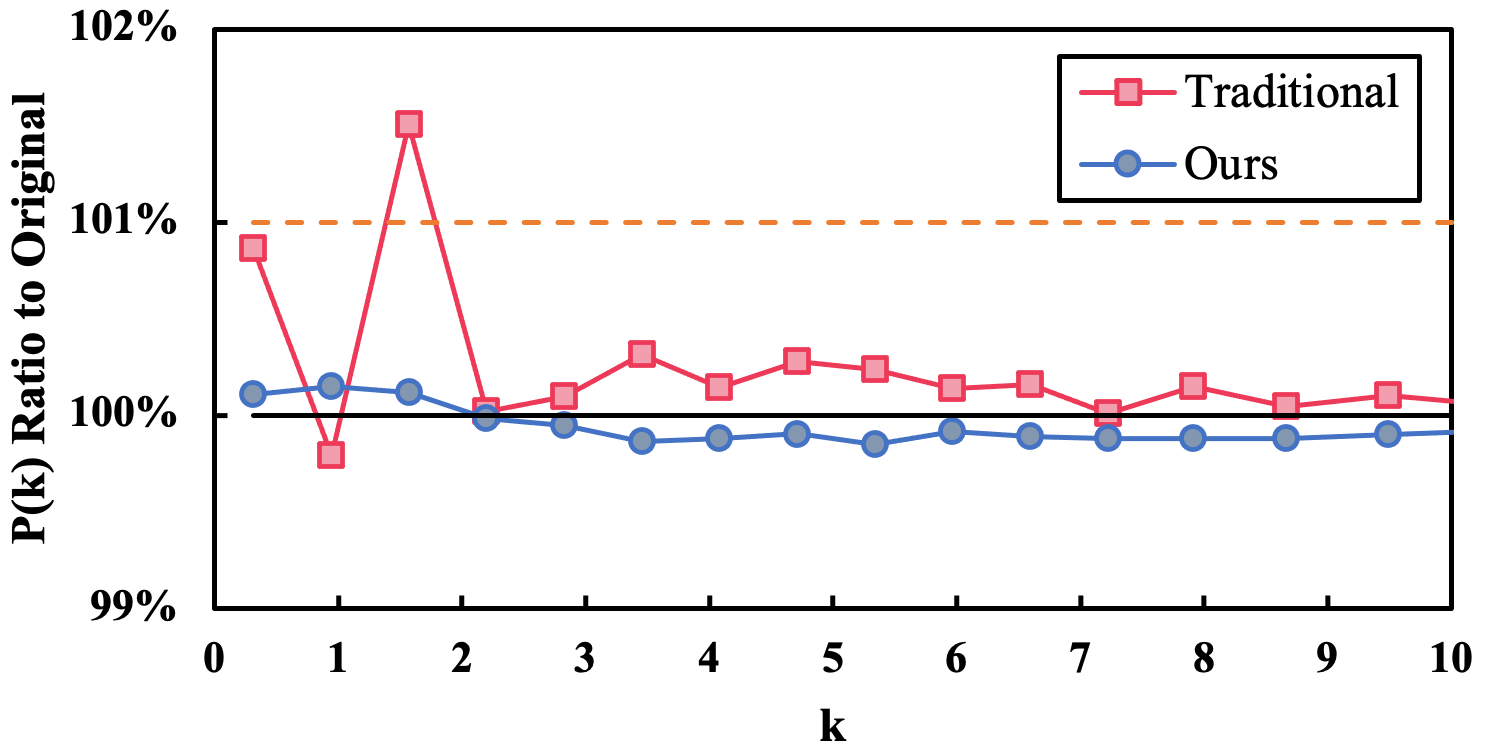}
    \vspace{-5mm}
    \caption{Power spectrum analysis on Nyx's \texttt{baryon density} field. Black solid line is the power spectrum on the original data for reference. Orange dashed line is the upper limit of acceptable power spectrum on the reconstructed data.}
    \label{fig:fig-4-3}
\end{figure}

When considering power spectrum as post-hoc analysis, which is the case for all data fields other than \texttt{baryon density} in our tested dataset, we generally require $P(k)$ ratio of reconstructed data to original data to keep within a $\pm1\%$ for $k < 10$, so that the simulation results can be compared and verified by our cosmological observation capability. Figure~\ref{fig:fig-4-3} demonstrates a power spectrum analysis with \texttt{baryon density} from a Nyx dataset. Note that one data point from data compressed by the traditional method has exceeded the acceptable error range. Our strategy, however, successfully bounds the analysis error within the acceptable error range without expensive trial-and-error. Note that in our experiment we choose $2\sigma$ from Equation~\ref{equ-4} mapped to an acceptable error range. Based on our model, this can provide a 95.4\% of confidence for no escaping error. In practice, this decision can be changed by potentially lower overall compression ratio for higher post-hoc analysis error control.

\begin{figure}[]
    \centering
    \vspace{-6mm}
    \includegraphics[width=0.9\linewidth]{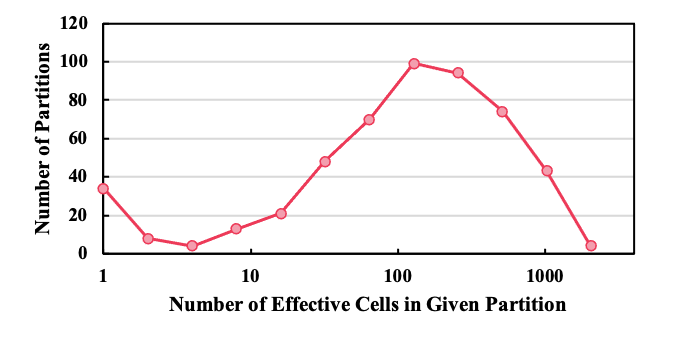}
    \vspace{-6mm}
    \caption{Histogram of effective cell count from all $512$ data partitions of $1024\times1024\times1024$ \texttt{baryon density} data.}
    \label{fig:fig-4-8}
    \vspace{-1mm}
\end{figure}

For the baryon density field, not only does it use power spectrum as post-hoc analysis, halo finder is also used in this field as an important workflow. From our proposed model in Section~\ref{sec:analysis}, we extract the number of cells with a value range between $(t_\text{boundary}-eb, t_\text{boundary}+eb)$ in each partition as features, since their values need to be preserved for halo finder. Figure~\ref{fig:fig-4-8} shows the number of such effective cells varies greatly for given partition, consider x-axis is log scaled for information demonstration. A dispersed histogram means a high potential of preserving features for some partitions while sacrificing features from others to achieve high compression efficiency. Note that here we do not need to re-determine this number every time we select a new error bound for the partition, because in small ranges the value distribution can be considered as evenly distributed as discussed in Section~\ref{sec:analysis}. We only need to extract this number once with $eb = 1.0$ (a fairly high error-bound setting) and get eb-cell function $n_{bc} = n \times eb$, where $n$ is the number of effective cells found with $eb = 1.0$. We also note that such effective cells also have a correlation to mean value, usually get more $n_{bc}$ from partitions with higher mean values. This illustrates a competition between selecting error bound based on compressibility or based on feature density. However, in our evaluation, when optimizing based on the quality-ratio modeling for FFT-based analysis, halo finder result can also be satisfied so that the RMSE of alternated halo mass and original halo mass remains $1\pm0.01$. Unless a higher accuracy is needed by halo finder, we use the optimization strategy discussed in Section~\ref{sec:analysis}, which can provide 29.8\% higher accuracy in terms of halo mass error, compared to the traditional method.

\begin{figure}[]
    \centering
    \includegraphics[width=0.95\linewidth]{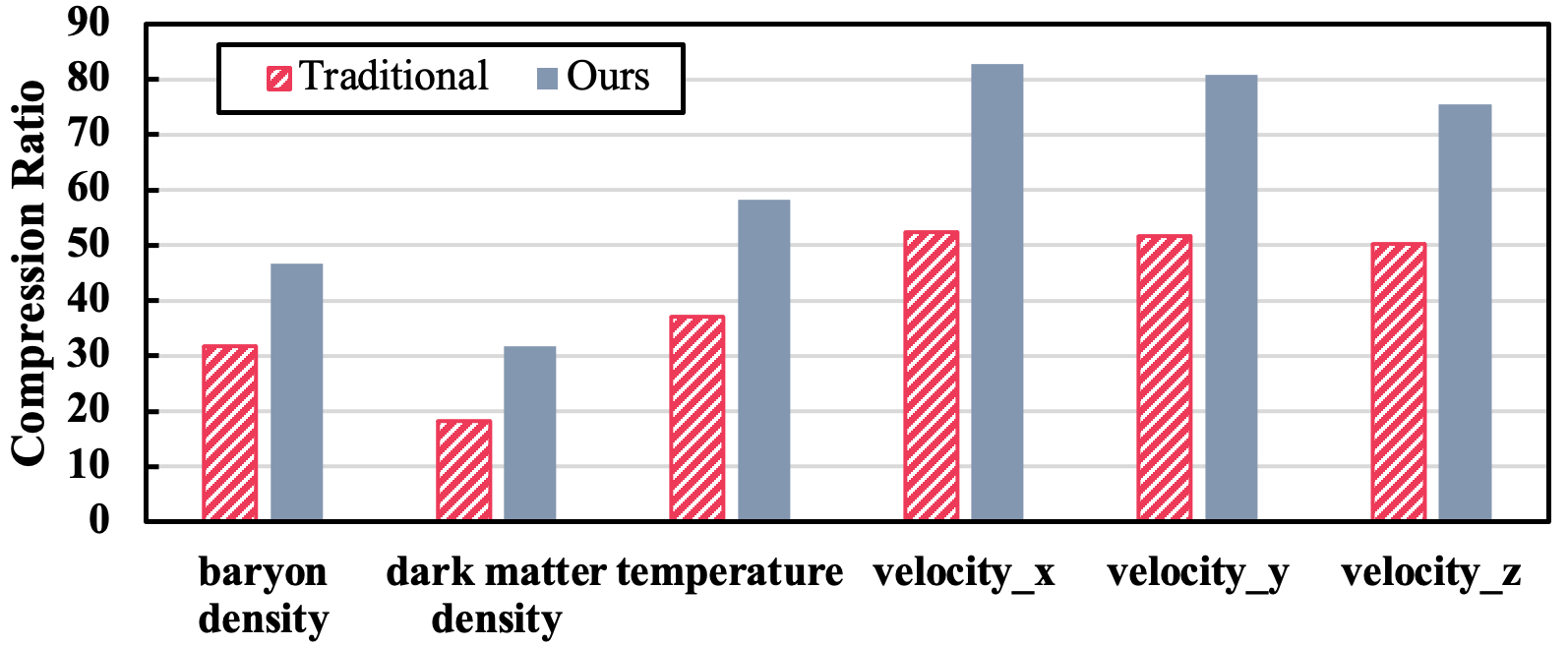}
    \vspace{-4mm}
    \caption{Compression ratio comparison between our and traditional methods on all 6 Nyx fields. All the reconstructed data satisfy the quality requirement of post-hoc analysis.}
    \label{fig:fig-4-4}
\end{figure}

\begin{figure}[]
    \centering
    \vspace{-8mm}
    \includegraphics[width=0.95\linewidth]{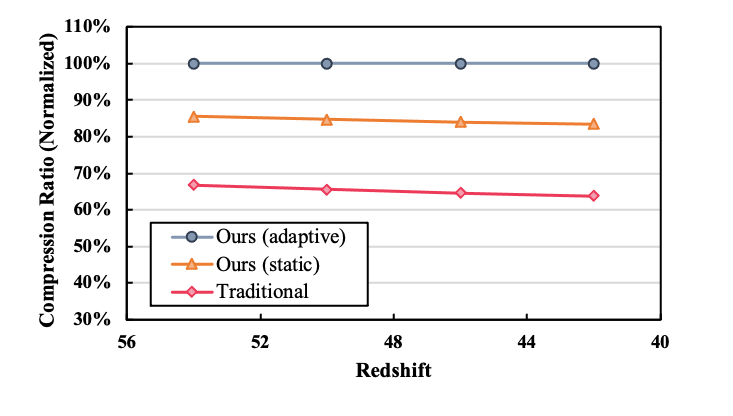}
    \vspace{-8mm}
    \caption{Compression ratio comparison between our and traditional methods on multiple redshifts' data using \texttt{baryon density} field. Compression ratio is normalized to our optimized ratio.}
    \label{fig:fig-4-5}
    \vspace{-1mm}
\end{figure}

Figure~\ref{fig:fig-4-4} shows the compression ratio improvement of our fine-grained adaptive lossy compression solution over the traditional method on the tested dataset while satisfying the data quality requirement for post-hoc analysis. Considering all 6 fields in the tested Nyx dataset, our solution provides an improvement of compression ratio by 56.0\% on average. The main advantage of our proposed technique is a more precise error-bound control for every partitions. On the other hand, our solution also requires no trial-and-error effort thanks to our specifically designed post-hoc analysis error modeling, %
which can provide a \textsc{near-optimal} solution with little optimization overhead imposed. The traditional method traverses all possible error bounds for a given dataset and can only provide a sub-optimal solution to maintain the high post-hoc analysis quality. 
For example, velocity data in Nyx simulation are highly random \cite{tao2017exploration,jin2020understanding}, thus simply deploying different error bounds to different partitions will not provide a very high compression efficiency.
The improvement on velocity data shown in Figure~\ref{fig:fig-4-4} mainly because of the highly accurate error-bound estimation. 
In fact, in order to guarantee the unpredictable post-hoc analysis error within acceptable for multiple snapshots, simulation users usually choose a relatively lower error-bound for lossy compressor based on empirical studies~\cite{jin2020understanding} compared to the optimized solution. 

In Figure~\ref{fig:fig-4-5}, we show the performance of our solution with various redshifts evolved to reduce over time (i.e., from 54 to 42 in our tested data). The static implementation of our solution (in yellow) is 
to optimize the error bounds for all partitions once at the early stage of simulation when the redshift is lower and keep using them for all following snapshots. 
We observe that the static implementation impacts the compression ratio to some extent, since the simulation data evolve to a different state and require an adjustment of error-bound combination for the highest compression efficiency. 

\begin{figure}[]
    \centering
    \includegraphics[width=0.95\linewidth]{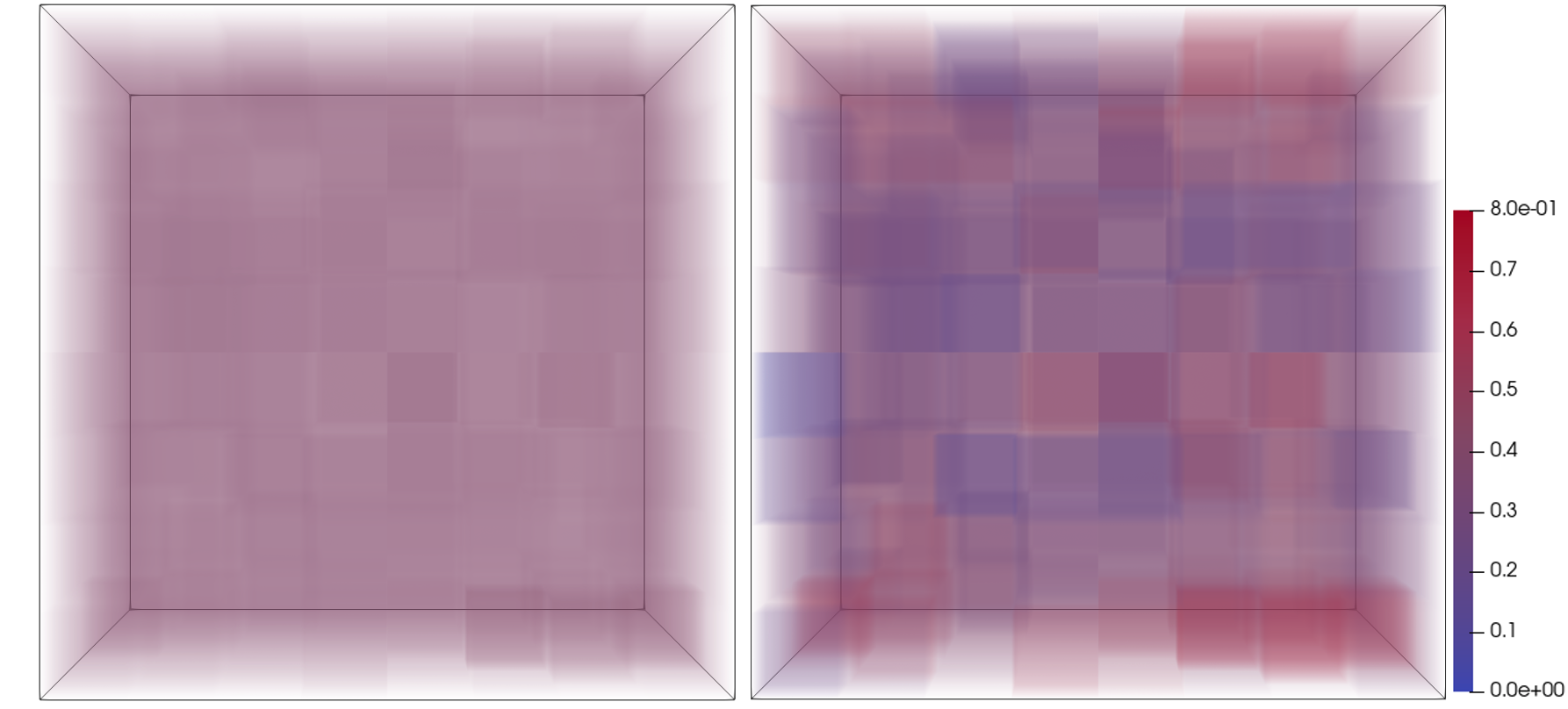}
    \vspace{-4mm}
    \caption{Comparison of our optimized error bounds on the data with larger redshift (left, early in simulation) and the data with lower redshift (right, late in simulation).}
    \label{fig:fig-4-10}
\end{figure}

Figure~\ref{fig:fig-4-10} illustrates the difference of error-bound configurations optimized by the early-stage dataset and by the redshift-42 dataset. Most of the data partitions in the early stage with larger redshift are smooth and close to each other, resulting in similar optimized error bounds for all partitions.
Compared to the traditional method, our solution can provide a consistent improvement on multiple snapshots. Note the improvement increases slightly as the redshift decreases. This is because cosmological simulation evolves to sparser formation, which increases the feature density or compressibility difference between partitions and can thus benefits from our proposed solution.

\begin{figure}[]
    \centering
    \vspace{-4mm}
    \includegraphics[width=0.9\linewidth]{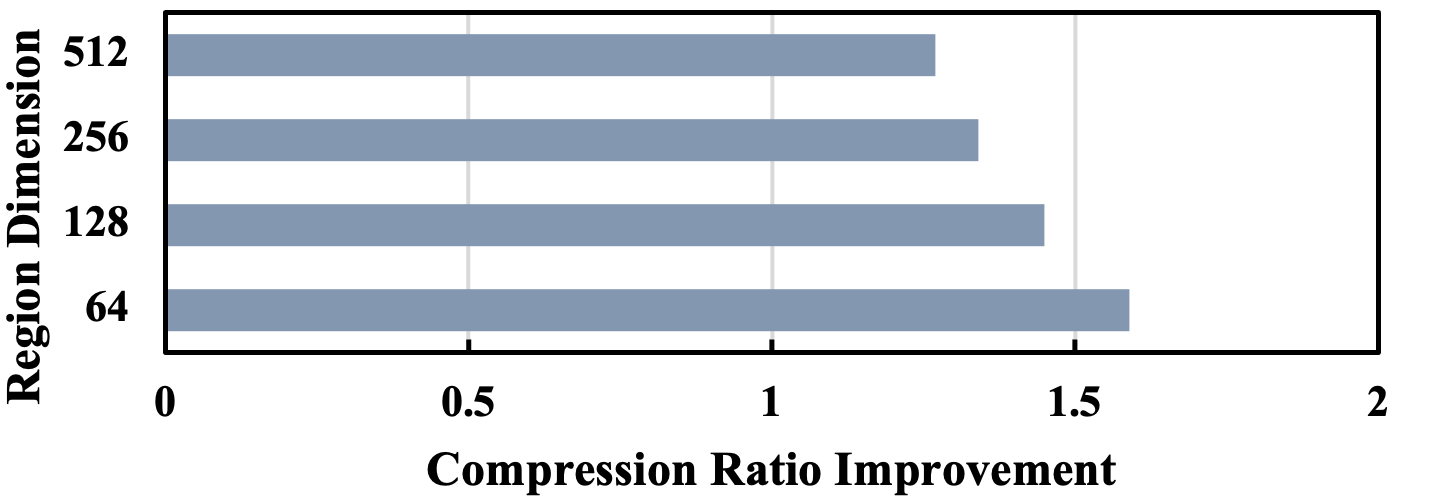}
    \vspace{-4mm}
    \caption{Compression ratio improvement with different partition sizes (compared to traditional method).}
    \label{fig:fig-4-6}
\end{figure}

\begin{figure}[]
    \centering
    \includegraphics[width=0.9\linewidth]{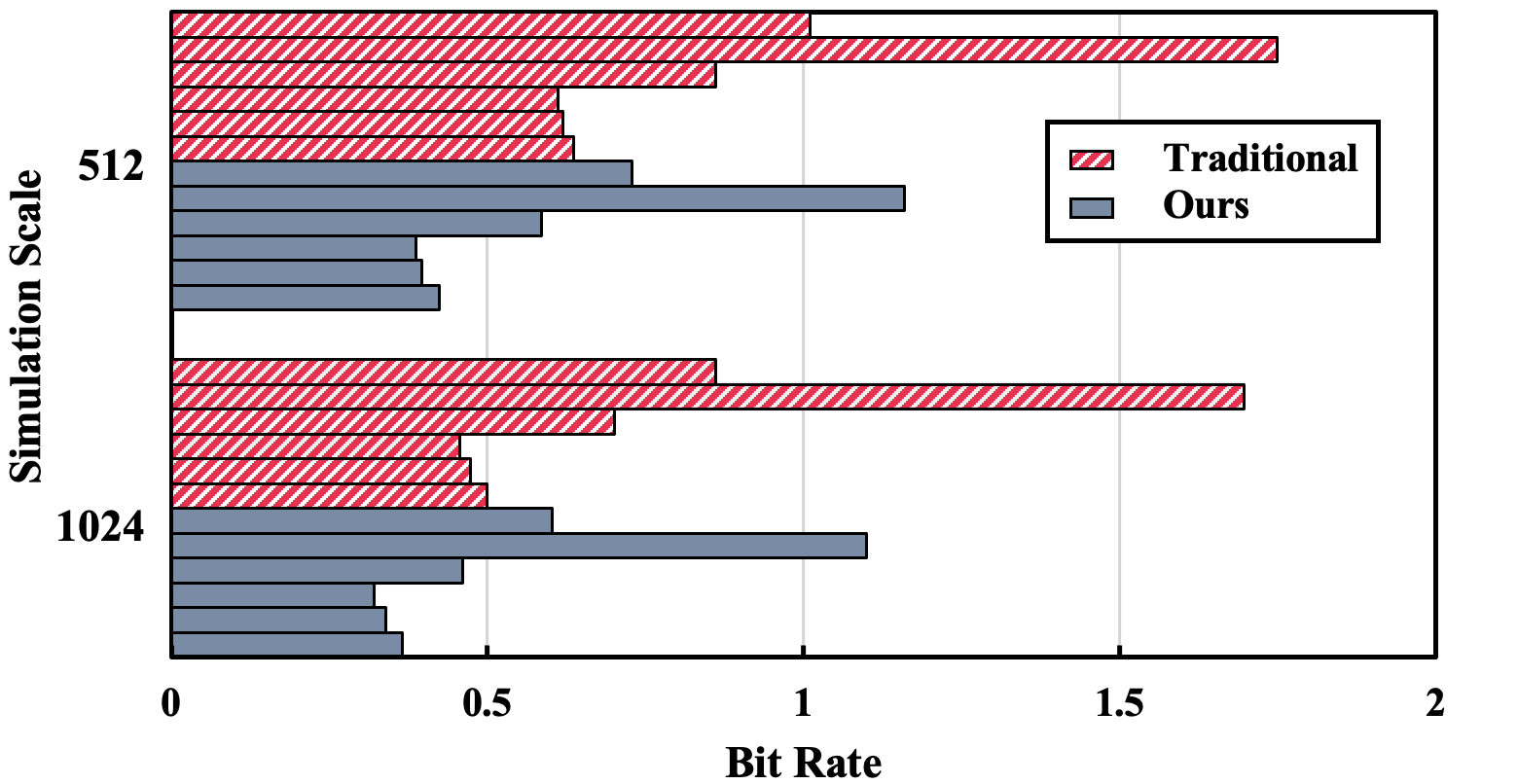}
    \vspace{-4mm}
    \caption{Compression efficiency improvement with different data sizes. Simulation scale is one dimension size of snapshot.}
    \label{fig:fig-4-7}
\end{figure}

We also evaluate our solution with different partition sizes and show the result in Figure~\ref{fig:fig-4-6}. We can observe that the overall compression ratio improvement increases as the partition size decreases, from 27.1\% to 56.0\% for partition dimension from 512 to 64, respectively. This is because larger partition size averages out the quality-ratio difference between partitions, which leads to a lower gain of compression efficiency from separately configuring error bounds to partitions, but it relies on a more accurate estimation of optimized error bounds. We suggest lowering the size of partition for higher utilization of our solution. Lastly, we evaluate our solution on the Nyx dataset with different simulation scales, shown in Figure~\ref{fig:fig-4-7}. It illustrates that our solution provides a consistent improvement over the traditional method on different simulation scales. More specifically, it provides average compression ratio improvement compared to the traditional method by 56.0\% and 51.9\% for simulation scale of 512 and 1024, respectively.

\subsection{Performance Evaluation and Discussion}

Traditional methods find desirable error bounds for a given dataset by traversing all possible candidates. With a smooth curve of ratio to error-bound for SZ lossy compression, a statically optimized error bound for the entire dataset can be always found by the trial-and-error process if there is no time limit. However, this is challenging for scientific datasets where quick-to-compute general distortion metrics cannot represent data quality for post-hoc analysis purposes~\cite{jin2020understanding}, and each trail requires compression, decompression, and post-hoc analysis. Thus, this method is not scalable and cannot be applied to extreme-scale simulations. Not to mention if pursuing higher compression efficiency by varying configuration for individual partition, the traditional trial-and-error method is simply infeasible considering the combination possibility of different partitions (e.g., $512$). In comparison, our solution only requires a very low overhead for the error-bound combination optimization with our theoretical models. As a result, our solution can be adapted to all snapshots regardless of the simulation scale. For FFT-based distortion metrics used for other scientific simulations, we have provided solutions in Section~\ref{sec:analysis}. 
For other types of post-hoc analysis, we propose to study the information necessary and build a specific model based on it for feature preserving, similar to our modeling methodology for halo finder.

In terms of the in situ overhead of our proposed solution, it only requires the mean value of each partition and the overall mean value of the dataset as parameters for all data fields.
To compute the mean value of each partition, it only takes about $1$$\sim$$1.5\%$ of the compression time overhead using our tested CPUs, while it takes almost no overhead on the tested GPUs. 
The overall mean value can be gathered by \verb+MPI_Allreduce+ after each partition generates its own mean value, which is almost negligible compared to compression throughput (e.g., 31.6 GB/s on a V100 GPU with cuSZ \cite{tian2020cusz}), not to mention both \texttt{baryon density} $(\rho_b)$ and dark matter density $(\rho_d{}_m)$ that require no \verb+MPI_Allreduce+ operation due to their fixed overall mean value set by the simulation. 
Moreover, we also need the number of effective cells as an extra parameter for \texttt{baryon density}. This process takes an extra time overhead of up to 5\% of compression time in our experiment. 
Overall, all the overheads introduced by our optimization are negligible comparing to the time of cosmological simulation or compression.
Therefore, our solution introduces very little overhead and provides a high compression efficiency improvement.
\section{Conclusion and Future Work}
\label{sec:conclusion}

In this paper, we propose to adaptively configure lossy compression in data partitions for cosmological simulation with newly designed rate-quality models for post-hoc analysis. We first propose an accurate model to estimate the post-analysis result considering lossy compression error, as well as a compression-ratio model based on theoretical analysis and experimental evaluation. We then further develop an adaptive optimization solution based on our models to dynamically determine the optimal compression parameters for different partitions instead of static selection, maximizing compression efficiency while offering an error control to post-hoc analysis. Our proposed fine-grained approach further improves the compression ratio by up to 73\% on the tested dataset compared to existing methods, making the compression ratio reach to $27.6$$\sim$$82.8\times$, with a very low time overhead of only 1\%. Our future work is 
to further apply our approach to other HPC applications and post-hoc analysis metrics such as climate simulation with SSIM.
\section*{Acknowledgments}

\small 
This work has been authored by employees of Triad National Security, LLC which operates Los Alamos National Laboratory under Contract No. 89233218CNA000001 with the U.S. Department of Energy/National Nuclear Security Administration.
This research was supported by the Exasky Exascale Computing Project (17-SC-20-SC), a collaborative effort of the U.S. Department of Energy Office of Science and the National Nuclear Security Administration. This research was supported by the U.S. National Science Foundation under Grants OAC-2034169 and OAC-2042084. We would like to thank NERSC for providing access to the Cori supercomputer and TACC for providing access to the Frontera supercomputer. We would like to thank the NYX team at Lawrence Berkeley National Laboratory for granting us access to cosmology datasets.

\bibliographystyle{ACM-Reference-Format}
\bibliography{refs.bib}

\end{document}